\documentclass[twocolumn,superscriptaddress,floatfix,prl,aps]{revtex4-1}
\usepackage{graphicx,amsfonts,amssymb,amsmath,amssymb,hyperref,hypcap,enumerate}
\usepackage{xcolor}
\usepackage{soul}
\usepackage{newtxtext,newtxmath} 
\usepackage{scalerel}
\usepackage{multirow}
\newcommand{\be}{\begin{equation}}
\newcommand{\ee}{\end{equation}}
\newcommand{\bmat}{\begin{pmatrix}}
\newcommand{\emat}{\end{pmatrix}}

\newcommand{\mb}[1]{\mathbf{#1}}
\newcommand{\bs}[1]{\boldsymbol{#1}}
\newcommand{\la}{\langle}
\newcommand{\ra}{\rangle}

\newcommand{\vep}{\varepsilon}
\newcommand{\ep}{\epsilon}

\begin{document}

\title{Weak Coupling Theory of Magic-Angle Twisted Bilayer Graphene}
\author{Jihang Zhu}
\affiliation{Condensed Matter Theory Center and Joint Quantum Institute, Department of Physics, University of Maryland,
College Park, Maryland 20742, USA}
\affiliation{Max Planck Institute for the Physics of Complex Systems, 01187 Dresden, Germany}
\author{Iacopo Torre}
\affiliation{ICFO-Institut de Ci\`{e}ncies Fot\`{o}niques, The Barcelona Institute of Science and Technology, Av. Carl Friedrich Gauss 3, 08860 Castelldefels (Barcelona),~Spain}
\author{Marco Polini}
\affiliation{Dipartimento di Fisica dell'Universit\`a di Pisa, Largo Bruno Pontecorvo 3, I-56127 Pisa, Italy}
\affiliation{Istituto Italiano di Tecnologia, Graphene Labs, Via Morego 30, I-16163 Genova, Italy}
\affiliation{ICFO-Institut de Ci\`{e}ncies Fot\`{o}niques, The Barcelona Institute of Science and Technology, Av. Carl Friedrich Gauss 3, 08860 Castelldefels (Barcelona),~Spain}
\author{Allan H. MacDonald}
\affiliation{Physics Department, University of Texas at Austin, Austin TX 78712 USA}

\begin{abstract}
Strong correlations occur in magic-angle twisted bilayer graphene (MATBG) when the octet of flat moir\'e minibands centered on charge neutrality (CN) is partially occupied.  The octet consists of a single valence band and a single conduction band for each of four 
degenerate spin-valley flavors.
Motivated by the importance of Hartree electrostatic interactions 
in determining the filling-factor dependent band structure, we use a 
time-dependent Hartree approximation to gain insight into electronic 
correlations.  We find that the electronic compressibility is dominated 
by Hartree interactions, that paramagnetic states are stable over a range of density near CN, and 
that the dependence of energy on flavor polarization is strongly overestimated by mean-field theory.   
\end{abstract}

\maketitle


{\em Introduction---}
The energy bands of twisted bilayer graphene (TBG) have a four-fold spin-valley flavor degeneracy.
As a magic twist angle near $\theta = 1^{\circ}$ is approached,
the two sets of four-fold degenerate bands closest to the neutral system Fermi energy
approach each other and narrow \cite{BM}, converting graphene from a weakly-correlated Fermi liquid 
to a strongly correlated system \cite{cao2018unconventional,cao2018correlated,andrei2020graphene,balents2020superconductivity} 
with a rich variety of competing states,
including superconductors, insulating flavor ferromagnets, and metallic flavor ferromagnets.
The ferromagnetism is 
reminiscent of but distinct from that exhibited by Bernal-stacked bilayer graphene in the 
quantum Hall regime \cite{McCann2006, Barlas2008, Cote2010, Barlas2012, Feldman2009, Martin2010, Weitz2010, Jung2011} and is now clearly established \cite{balents2020superconductivity,cao2018correlated, serlin2020intrinsic, sharpe2019emergent, stepanov2020competing,lu2019superconductors,polshyn2020electrical,zondiner2020cascade,wong2020cascade,das2020symmetry, khalaf2021charged,saito2020isospin,xie2021fractional,wu2021chern,nuckolls2020strongly,kang2019strong,lian2021twisted,liu2022visualizing,wang2020correlated,wang2022kekul,Zhu2020TBG}
as a prominent part of the physics of MATBG.
In contrast to the quantum Hall case, in which eight Landau bands are filled sequentially to minimize the exchange energy,
MATBG ground states appear \footnote{These statements are based on a minimal interpretation of measurements of electronic 
compressibility \cite{zondiner2020cascade}, 
tunneling spectroscopy \cite{wong2020cascade} and 
weak-field Hall effects \cite{xie2020weak}.  
Other scenarios in which flavor symmetries are broken even at CN cannot be ruled out at present and are 
assumed in some theoretical work. Note that the cascades observed in the high-temperature compressibility experiment \cite{saito2020isospin} (up to 100 K) are not indicative of ground state phase transitions. These high-temperature cascades may result from fluctuations of local moments, which can be explained by localized orbitals and the heavy fermion model \cite{song2022magic, JKang_heavyfermion_2021}.} 
not to have any broken symmetries for a range of filling factors near 
CN, and in broken symmetry states 
to keep the filling factors of partially occupied flavors 
$\nu_f$ inside an interval $(-\nu_h^*,\nu_e^*)$, 
where $\nu_h^*$ and $\nu_e^*$ are maximum hole and electron filling factors.
($\nu_f\equiv (N_f-M)/M$ where $N_f$ is the number of flat band electrons with flavor $f$ and $M$ is the 
number of moir\'e cells in the system; $\nu=\sum_f \nu_f$.)

In this Letter, we address some unusual aspects of the correlation physics of MATBG from the weak-coupling point of 
view (one shot GW approximation).
We find that the average compressibility is dominated 
by Hartree interactions, that unbroken symmetry states are stable over a range of density near CN, and 
that the dependence of energy on flavor polarization is strongly overestimated by mean-field theory.   
Below we first explain the technical details of our calculations and then discuss the 
relationship of our findings to those obtained using other approaches to MATBG interaction physics.


{\em Moiré-Band Weak-coupling Theory---}  
The one shot GW approximation, also known as the random phase 
approximation (RPA), is a perturbative method that accounts for 
dynamic screening of long-range Coulomb interactions. 
It is commonly used \cite{golze2019gw,deslippe2012berkeleygw} in {\it ab initio}
electronic structure theory to understand collective electronic behaviors, especially as probed by optical or photoemission spectroscopy.
Although rigorously justified \cite{gell1957correlation} only in 
weakly interacting systems, it has recently attracted interest
\cite{olsen2013random} as a universal and accurate method for total energy calculations in 
many real materials, including \cite{olsen2017assessing} strongly correlated Mott insulators.

\begin{figure*}[!t]
\includegraphics[width=2.0\columnwidth]{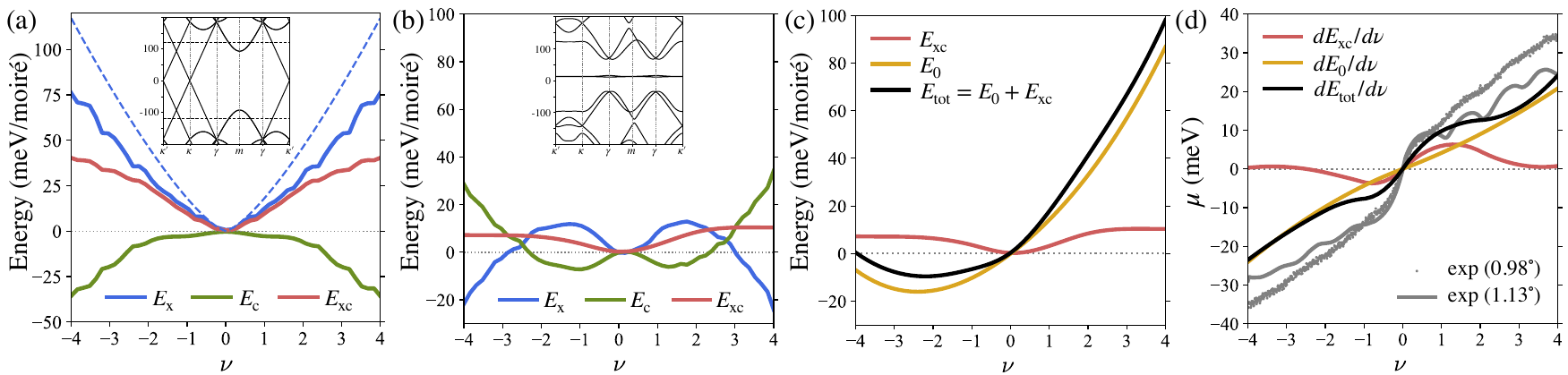}
  \vspace{-10pt}
  \caption{\small Energies of paramagnetic states as a function of $\nu \in [-4,4]$ for (a) a decoupled-bilayer and (b-d) $1.1^\circ$-TBG.
  (a-b) Exchange ($E_x$) and RPA correlation ($E_c$) energies as defined in Eqs.~(\ref{Eq:Ex}-\ref{Eq:Ec}).
  The insets show the corresponding single-particle band structures. The black dashed lines in the inset of (a) mark the Fermi level
  for $\nu=\pm 4$. The blue dashed line in (a) is the
 exchange energy calculated using the approximate analytical expression Eq.~(\ref{Eq:Ex_analy}). 
 (c) The SCH energy $E_0$ \cite{SeeSM} and the RPA total energy $E_{\text{tot}}$. (d) The calculated chemical potential $\mu = dE_{\text{tot}}/d\nu$ with its zero shifted to the chemical potential at $\nu=0$. The grey dots ($0.98^\circ$) and the grey line ($1.13^\circ$)
 plot measured chemical potentials from Ref.\cite{zondiner2020cascade}.
 All energies are given relative to CN with the zero of energy at the 
 neutral system Fermi level.
  }
  \label{Fig:Exc_unbroken}
\end{figure*}

\begin{figure*}
\includegraphics[width=1.7\columnwidth]{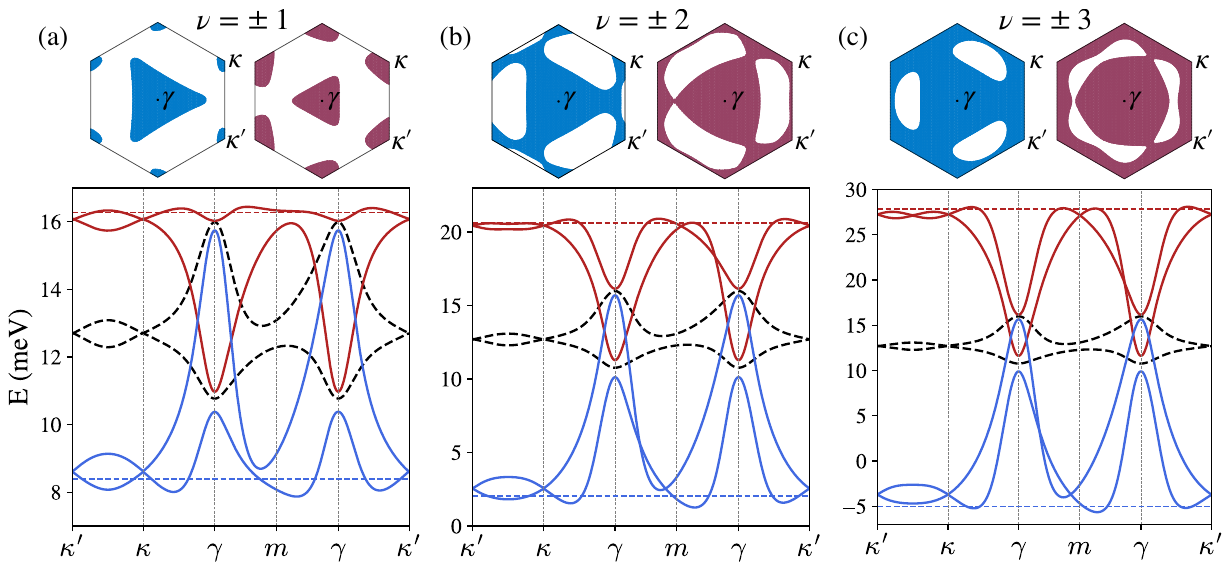}
  \vspace{-10pt}
  \caption{\small 
  The SCH paramagnetic state bands (colored lines) and the corresponding Fermi surfaces (shaded areas) at a series of 
$\nu$ values on hole-doped (blue) and electron-doped (red) sides.
  The black dashed line in each spectrum is the single-particle band structure and the colored dashed horizontal lines mark Fermi levels.
  At $\nu=-3$, the flat valence band is $1/4$ full and the occupied states are those whose charge density is 
  most peaked near minima of the external potential produced by remote band charges.  At $\nu=-1$, the flat valence band is 
  at $3/4$ filling.  Holes in the valence band remain near $\gamma$, which would be the valence band bottom if 
  Hartree corrections were not included.  Holes near $\gamma$ are finally filled only around $\nu=-0.3$ (see Fig.~\ref{FigSM_FS} in SM~\ref{SM:scHartree}) as $\nu$ approaches zero and Hartree energies finally become small compared to band energies.  
  The Fermi surfaces at filling factors $+ \nu$ and $-\nu$ (for example $\nu=2$ and $\nu=-2$) would be identical for any $\nu$ if the model had exact particle-hole symmetry.  At filling factors away from $\nu=0$,
  the SCH band width is dominated by the Hartree mean-field contribution.
  }
  \label{Fig:SCH_FS}
\end{figure*}

In this Letter we employ RPA theory to approximate the dependence of energy on the 
total band filling factor and on the partitioning of electrons between the four spin-valley flavors of MATBG.
Because the number of electrons for each flavor is a good quantum number, 
we can approximate the magnetic energy landscape by adding exchange-correlation (xc) corrections $E_{\rm xc}$ to
the self-consistent Hartree (SCH) energies of flavor polarized states.  The RPA theory is motivated by the unusual
property of MATBG, illustrated in Fig.~\ref{Fig:SCH_FS} by plotting SCH bands at a 
series of band filling factors, that the band filling dependence of its total energy is dominated \cite{xie2020nature, Guinea_scHF_2018, Cea_scH_2019, Rademaker_scH_2019, Cea_scHF_2020}
by a Hartree mean-field contribution. The SCH energy increases rapidly as the flat bands are 
filled as shown in Fig.~\ref{Fig:Exc_unbroken}(c), and dominates the experimentally measured compressibilty.  
The RPA accounts both for this energy, and for dynamic fluctuation corrections to it.  

The xc correction to the SCH energy can be expressed \cite{SeeSM}, in terms
of a coupling-constant integral of the pair correlation function.  This quantity can in turn
be related to the density response function by
\be
\label{Eq:Exc_TBG}
\begin{split}
E_{\rm xc} 
&= \frac{1}{2} \sum\limits'_{\mb{q},\mb{g}} V_{\mb{q}+\mb{g}} \Big[ -\frac{1}{\pi } \int_0^1 d\lambda \int_0^\infty d\omega \; \chi^{\mb{g} \mb{g}}(\mb{q},i\omega;\lambda) -1 \Big],
\end{split}
\ee
where $V_{\mb{q}} = 2\pi e^2/q \epsilon_{_{\rm BN}}$ is the two-dimensional (2D)
Coulomb interaction accounting for hexagonal boron nitride (hBN) dielectric screening
with the dielectric constant chosen to be $\epsilon_{_{\rm BN}}=5.1$ throughout the paper, $\mb{q}$ is a wavevector in the moir\'e Brillouin zone (MBZ),
$\mb{g}$ is a moir\'e reciprocal lattice vector, and the 
prime on the sum excludes the $\mb{q}=\mb{g}=0$ term which
contributes only a gate-geometry-dependent constant \cite{SeeSM}.
In Eq.~(\ref{Eq:Exc_TBG}) $\chi^{\mb{g} \mb{g}}$ is a diagonal matrix element 
of the density response function, 
which is a matrix in reciprocal lattice vectors because of system's discrete translational symmetry,
and the frequency integration used to obtain equal time correlations has been rotated to the imaginary axis.

Equation~(\ref{Eq:Exc_TBG}) is formally exact.  In RPA (time-dependent Hartree) we replace $\chi$ in Eq.~(\ref{Eq:Exc_TBG})
by
\be
\label{chi_RPA}
\begin{split}
\chi(\lambda) 
&= \tilde{\chi}_{_{\rm H}}(1-\lambda V \tilde{\chi}_{_{\rm H}}
)^{-1} \\
&=\tilde{\chi}_{_{\rm H}} + \lambda \tilde{\chi}_{_{\rm H}} V\tilde{\chi}_{_{\rm H}}(1-\lambda V \tilde{\chi}_{_{\rm H}})^{-1},
\end{split}
\ee
where $\tilde{\chi}_{_{\rm H}}$ is the single-particle density response function 
calculated from the SCH bands \cite{Novelli_TBGoptical_2020},
summing over independent contributions from all four flavors:
\be
\label{Eq:chiH4f}
\tilde{\chi}_{_{\rm H}} = \sum\limits_{f=1}^4 \tilde{\chi}_{_{\rm H}}^f.
\ee
Possible improvements to this approximation are discussed later.

When inserted in Eq.~(\ref{Eq:Exc_TBG}), the second form for the right-hand-side of 
Eq.~(\ref{chi_RPA}) separates the exchange energy $E_{\rm x}$, the contribution that is first order in $V$,
from the full fluctuation correction $E_{\rm xc}\equiv E_{\rm x}+E_{\rm c}$, allowing us to 
carefully account for its subtly convergent frequency integral.
After integrating over $\lambda$, the exchange energy can be rewritten in the standard Slater determinant form \cite{SeeSM}:  
\begin{widetext}
\be\label{Eq:Ex}
E_{\rm x} = -\frac{1}{2A} \sum\limits'_{\mb{q},\mb{g}} V_{\mb{q}+\mb{g}} \sum\limits_{\substack{f,\mb{k} ,\alpha \\ \beta, \mb{g}_1,\mb{g}_2}} \big[\delta \bar{\rho}^f(\mb{k}) + 2 \bar{\rho}^{0f}(\mb{k}) \big]_{\alpha,\mb{g}_1;\beta,\mb{g}_2} \delta \rho^f_{\alpha, \mb{g}_1+\mb{g};\beta,\mb{g}_2+\mb{g}}(\mb{k}+\mb{q}),
\ee
\end{widetext}
where $ \delta \rho^f(\mb{k})=
\sum\limits_n \Big( \hat{z}_{n}(\mb{k}) \hat{z}_{n}^\dagger(\mb{k}) \Theta_{n\mb{k}} - \hat{z}^0_n(\mb{k}) \hat{z}^{0\dagger}_n(\mb{k}) \Theta^0_{n\mb{k}} \Big)$ 
is the density matrix projected to flavor $f$ relative to that 
of a charge neutral decoupled bilayer, $\delta \bar{\rho}$ is the complex conjugate of 
the corresponding matrix element of $\delta \rho$, $\hat{z}_n(\mb{k})$ and $\hat{z}^0_n(\mb{k})$ are plane-wave representation SCH and 
neutral-decoupled-bilayer quasiparticle eigenvectors, and 
$\Theta_{n\mb{k}}$ and $\Theta^0_{n\mb{k}}$ are the corresponding occupation numbers.
In Eq.~(\ref{Eq:Ex}) $\mb{g}, \mb{g}_1, \mb{g}_2$ are moir\'e reciprocal lattice vectors,
$\mb{k}$ and $\mb{q}$ are momenta in MBZ, 
$\alpha$ and $\beta$ are layer and sublattice indices and $A$ is the area of the 2D system.
Because of their negative energy seas, continuum models of graphene multilayers  
are able to determine total energies only up to a reference energy (per area) that is 
a linear functions of electron density,
$\varepsilon_{\text{ref}}=\varepsilon_0+\mu_0 n$; Eq.~(\ref{Eq:Ex}) chooses the zero of energy 
$\varepsilon_0$ to be the energy per area of neutral decoupled bilayers and the zero of chemical potential 
$\mu_0$ to be the energy of states at the top of the decoupled bilayer valence band.
The integration over the coupling-constant $\lambda$ in Eq.~(\ref{Eq:Exc_TBG}) can be performed analytically
to yield the correlation energy \cite{SeeSM}
\be\label{Eq:Ec}
\begin{split}
E_{\rm c}
&= \frac{1}{2\pi} \sum\limits'_{\mb{q}}
\int_0^\infty d\omega \text{Tr} \Big[ \sqrt{\mb{V}} \tilde{\bs{\chi}}_{_{\rm H}} \sqrt{\mb{V}} + \ln (1- \sqrt{\mb{V}} \tilde{\bs{\chi}}_{_{\rm H}} \sqrt{\mb{V}}) \Big],
\end{split}
\ee
where $\mb{V}$ and $\tilde{\bs{\chi}}_{_{\rm H}}$
are matrices in reciprocal lattice vector with implicit $\mb{q}$ and $\omega$ dependences.
The correlation energy must be regularized by subtracting its value in unbroken symmetry states 
at CN; its contribution to the chemical potential at CN is close to zero
because the models we study have approximate particle-hole symmetry.


{\em Paramagnetic State Energy ---}
We interpret our numerical results for the band filling $\nu$ dependence of the MATBG paramagnetic ground state 
energy (Fig.~\ref{Fig:Exc_unbroken}(b,c)) by comparing them with results for the decoupled bilayer \cite{Barlas_G_2007} 
(Fig.~\ref{Fig:Exc_unbroken}(a)) calculated in exactly the same way.
In both cases the exchange energy is positive at small $|\nu|$ because of \cite{Barlas_G_2007} rapid changes in Bloch state
spinors near the Dirac point.  The blue dashed line in Fig.~\ref{Fig:Exc_unbroken}(a)
is the exchange energy of an eight-Dirac-cone model \cite{Barlas_G_2007}:
\be
\label{Eq:Ex_analy}
\begin{split}
E_{\rm x}^{\rm D} = & \frac{\alpha \hbar c}{24 \pi} \frac{g}{\epsilon_{_{\rm BN}}} k_{_{\rm F}}^3
\ln \Big( \frac{k_{\rm c}}{k_{_{\rm F}}}\Big) + \text{regular terms}, \\
\end{split}
\ee
where $g=8$ and $k_{_{\rm F}}=(4\pi n/g)^{1/2}$. 
The exchange energy of MATBG is smaller than that of decoupled bilayers
because of the dominant role of the Hartree potential in shaping 
occupied band states wavefunctions.  In contrast to the decoupled bilayer case,
MATBG correlation energies are low near CN, because that is where the phase space for low-energy particle-hole 
excitations within the flat band octet is the largest.  The correlation energy is 
highest near $|\nu|=4$ because the gaps between flat and remote bands suppress fluctuations.   
In our calculations there is a small particle-hole asymmetry in all properties,
including the exchange and correlation energies, because we include non-local interlayer tunneling corrections \cite{xie2020weak}
to the Bistritzer-MacDonald (BM) MATBG model \cite{BM, SeeSM}.

Because of the partial cancellation between exchange and correlation effects, discussed again below in connection with 
flavor ferromagnetism, the difference between MATBG and decoupled bilayers is dominated by the SCH energy \cite{SeeSM} plotted in Fig.~\ref{Fig:Exc_unbroken}(c). The SCH energy is calculated relative to 
its value at CN, and its slope at CN is finite because the bare flat bands are 
centered around $\varepsilon_{\text{fb}} \approx 12$ meV (see Fig.~\ref{Fig:SCH_FS}) in the non-local BM model we employ.
The chemical potential $\mu$, the energy 
to add a single-electron increases steadily as the flat bands are filled mainly because of Hartree effects.  
We find that the chemical potential difference between full and empty flat bands is $\sim 50$ meV.
When the bands are nearly empty, added electrons occupy 
regions in the moir\'e unit cell in which the mean-field potential from remote band electrons is most attractive.  
When the bands are nearly filled, on the other hand, it follows from approximate 
particle-hole symmetry that electrons 
occupy the same region but the Hartree mean-field potential is now repulsive.

\begin{figure*}
  \includegraphics[width=2.0\columnwidth]{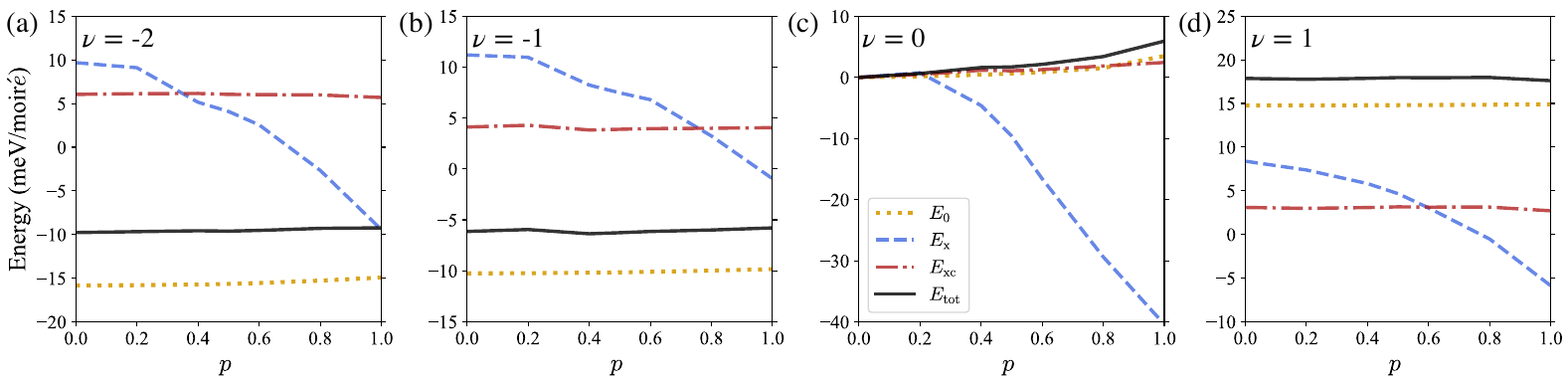}
  \vspace{-10pt}
  \caption{\small 
  SCH energy $E_0$ (yellow dotted lines), exchange energy $E_{\rm x}$ (blue dashed lines), exchange-correlation energy $E_{\rm xc}$ (red dash-dotted lines) and RPA total energy $E_{\rm tot}$ (black solid lines) as a function of 
  polarization $p$ at (a) $\nu=-2$, (b) $\nu=-1$, (c) $\nu=0$ and (d) $\nu=1$.  
  $p$ characterizes the degree of flavor polarization as explained in the main text.   
  $p=0$ corresponds to the paramagnetic state and $p=1$ corresponds to full flavor polarization.
  }
  \label{Fig:Exc_broken}
\end{figure*}

In Fig.~\ref{Fig:Exc_unbroken}(d) we compare our results for the 
filling factor dependence of the chemical potential across the full range of flat band filling
with experimental results published in Ref.~\cite{zondiner2020cascade}.  The total shift in chemical potential 
is somewhat larger in experiment than in theory.  Since the states near the full and empty flat band limit 
are not expected to be strongly correlated, we attribute this small discrepancy to weak
mixing between flat and remote bands and small inaccuracies in the
continuum model we employ. 
The most striking feature of these results is shared between theory and experiment,
namely that the chemical potential increases approximately linearly with band filling factor \cite{zondiner2020cascade,wong2020cascade,Tomarken_capacitance2019, scHF_Cea_2022}.  In MATBG experiments, 
structures do emerge at some filling factors that are thought to be due to first order flavor-symmetry breaking
phase transitions at low temperatures, which we now address, and at higher temperatures to surviving local moment fluctuations \footnote{One important limitation of the weak-coupling approach is that it does not capture the influence of unordered fluctuating local moments.}.

{\em Flat Band Flavor Ferromagnetism ---}
The RPA energy calculation can be carried out for any set of flavor-dependent filling factors.  
Typical numerical results \footnote{A more complete set of numerical results 
for a wide variety of filling factors and flavor polarizations are gathered in SM~\ref{SM:Table_energy} Table~\ref{tab:energy}.} are summarized in Fig.~\ref{Fig:Exc_broken}.  
The $\nu=0$ polarized states in Fig.~\ref{Fig:Exc_broken}(c) have  
filling factor $p$ for two flavors and filling factor $-p$ for the other two flavors.  Increasing $p$ shifts states from the valence bands of two flavors to the conduction bands of the other two flavors. 
Because of MATBG's approximate particle-hole symmetry, this polarization path does not strongly influence the charge density, which remains approximately uniform at this filling factor for all values of $p$, as illustrated in Figs.~\ref{Fig:rhor_vs_p},\ref{Fig:rhor_vs_nu}.
The main point to notice is that fully polarized states are strongly favored by exchange energies, but this 
energy gain is almost perfectly cancelled by the correlation energy which strongly favors states in which each flavor 
is half filled.  Similar results are obtained at other filling factors. 
The family of polarized states at $\nu=-2$ in Fig.~\ref{Fig:Exc_broken}(a) have filling factor $-(1+p)/2$ for two flavors 
and filling factor $-(1-p)/2$ for the other two flavors; increasing $p$ shifts electrons between
valence bands with different flavors and the charge density is non-uniform at all values of $p$.
For $\nu=\pm 1$, the flavor polarization path illustrated in Fig.~\ref{Fig:Exc_broken}(b,d) is 
$\nu=\pm (1+3p)/4$ for one flavor and $\nu=\pm (1-p)/4$ for the remaining three flavors. The exchange energy gain upon polarization is again almost exactly
cancelled by correlation, underscoring the dominance of the SCH energy. 
Once correlations are included the dependence of the SCH energy on $p$,
which was judged to be insignificant in previous self-consistent Hartree-Fock \cite{Liu_scHF_2021} calculations,
retains a role in the energy competition among different polarized states.

Within the RPA theory the cancellation between exchange and correlation for the polarization $p$
dependence of the energy can be understood in terms of Eqs.~(\ref{Eq:Exc_TBG}-\ref{chi_RPA}). 
The $p$ dependence of energy follows from that of $\tilde{\chi}_{_{\rm H}}$,
and this lies mainly in the range of low-frequency fluctuations within the flat band where 
the important matrix elements of $V \tilde{\chi}_{_{\rm H}}$ are much larger than 1 so that
$\chi(\lambda) \to V^{-1}$ (perfect screening), and the dependence of $E_{\rm xc}$ on polarization is lost.
Physically, correlations are already strong even in the paramagnetic state 
and there is little left to gain by flavor ordering.
Generally speaking, we find that the tendency toward flavor symmetry breaking is stronger 
at larger $|\nu|$ and stronger at positive $\nu$ than at negative $\nu$, as 
summarized in Fig.~\ref{Fig:Exc_polarize}, in agreement with most  
experiments \cite{zondiner2020cascade,Tomarken_capacitance2019,yankowitz2019tuning, stepanov2020competing}.
In addition we find that the difference in energy between polarized and paramagnetic states is drastically 
reduced by correlations from $\sim 40$ meV per moir\'e period to less than $\sim 3$ meV (Fig.~\ref{Fig:Exc_broken}(c)).


In MATBG broken $C_2T$ symmetry opens up a gap between the conduction and valence bands.  This type of broken 
symmetry within flavors is therefore common in mean-field calculations.  In our RPA calculations we find,
as summarized in Table~\ref{tab:energy_woC2T}, that
when $C_2T$ symmetry is broken by adding a sublattice-dependent potential of the type produced by 
aligned hBN substrates, flavor ferromagnetism is favored at almost all filling factors 
including those proximate to CN. This finding aligns well with experimental evidence suggesting 
that hBN alignment tends to favor states with broken symmetries \cite{sharpe2019emergent, serlin2020intrinsic, sharpe2021Orbital}, including 
quantum anomalous Hall states at fractional flat band fillings \cite{ serlin2020intrinsic}.

{\em The Magic-Angle Correlation Problem---}
In this Letter we have reported on the first RPA calculation for MATBG.
The RPA weak-coupling approach has the advantage that it accounts for dynamic screening of long-range
Coulomb interactions, but is less reliable than some other methods in accounting for short-distance correlations.
Competing methods often require tight-binding models, which in the case of MATBG 
have the disadvantage that they require the introduction of 
additional bands \cite{HCPo_TBG_TBmodel_2019} to compensate for fragile topology inherited from the isolated layer Dirac cones.
Our theory establishes the
crucial influence of correlations in
compressible metallic states in expanding
unbroken symmetry regions
in the MATBG phase diagram. The
RPA weak-coupling approach is also relevant for other
moir\'e materials that exhibit strong
correlations.

Our calculations include 146 remote valence and conduction bands per spin and flavor.
Our calclations are consistent with experimental indications that flavor ferromagnetism 
is common in both insulating and metallic states when the MATBG flat bands are partially filled, less likely close to CN, and 
more likely at positive filling factors than at negative filling factors.
The exchange energy gains that favor broken symmetry insulating ground states at integer $\nu$,
are comparable in size to correlation energy gains in closely competing metallic states with fewer or no broken symmetries.  
The resulting weak dependence of energy on magnetic state is consistent with small collective excitation energies of 
insulating states \cite{khalaf2020soft,kumar2020lattice} and with strong coupling 
approaches \cite{khalaf2021charged,song2022magic} that can 
be applied close to integer band fillings.  
Our calculations demonstrate that \cite{SeeSM} fluctuations in remote bands do not generally play a central role 
in MATBG properties except in the cases of nearly empty and nearly full bands.
This finding justifies the flat-band projection that is required to make 
non-perturbative finite-size numerical calculations \cite{potasz2021exact, Repellin_TBG_ED_2020, Xie_TBG_ED_2021, Soejima_TBG_DMRG_2020} feasible.  Perturbative calculations are approximate, but have the 
advantage that finite-size effects can be eliminated by taking dense momentum space grids; our 
calculations employ 432 $k$-points in the MBZ. 

Our calculation results can be compared directly to experimental results for 
the chemical potential $\mu$, which increases by $\sim 50$ meV as the flat 
bands are filled.  This compares to a dependence of energy on flavor polarization 
that is typically $\sim 3$ meV per moir\'e cell.
The positive compressibility we find, in agreement with experiment,
for MATBG electrons contrasts with the well-known negative compressibility of 
strongly interacting two-dimensional electron gas systems \cite{eisenstein1992negative, QuantumCapaGraphene_2013}, and 
is associated with unusual properties of the projected flat-band Hilbert space.  In MATBG models with exact particle-hole symmetry, the flat conduction and valence bands at the Fermi energy 
spatial structure within the moir\'e unit cell that precisely complements the total 
density of remote occupied bands, so that the total density is uniform.  The increase in chemical 
potential with filling factor is associated with the property that the 
non-uniform density of the remote bands is first eliminated and then restored with the opposite sign 
as the flat bands are filled.  We emphasize that unlike most calculations in the literature, 
which overstate dielectric screening to suppress interaction scales, all our results are obtained using 
a physically realistic hBN dielectric constant $\epsilon_{_{\rm BN}}=5.1$ \cite{SeeSM}.

The MATBG correlation problem is extraordinarily challenging and the RPA theory, like other 
approaches, has limitations.  Even though the flat band eigenstates have weak dispersion, their wavefunctions 
vary in a complex way across the MBZ. For this reason there is no simple Hubbard-like 
lattice model representation of the correlation problem.  Aside from the fascinating low-temperature 
superconducting instability, two key higher energy issues still do not have definitive answers.
i) What is the ground state at CN?  Is it the $p=0$ state of Fig.~\ref{Fig:Exc_broken},
which has no broken symmetries and strong correlations, or the $p=1$ state, which is a single
Slater determinant with analytically calculable excitations when remote band fluctuations are neglected?
ii) What is the Fermi surface in the range of filling factors surrounding $\nu=0$?  Is it the 
$\gamma$ centered Fermi surface of the $p=1$ state or the $\kappa,\kappa'$ centered Fermi surface of the 
$p=0$ state?  In either case how does the Fermi surface, at least as indicated by weak-field Hall 
measurements \cite{xie2020weak}, manage to avoid Liftshitz transitions over such a broad range of filling factors
$-1.8 \lesssim \nu \lesssim 0.9$ surrounding $\nu=0$?
For the first question we do not consider the weak-coupling answer (that $p=0$ is favored) to be definitive, but it 
certainly demonstrates that the two states are competitive.  The second question is especially troublesome if
one imagines that the ground state near $\nu=0$ is a doped $p=1$ state in which the band degeneracies have been
reduced from four to two and Fermi surface areas must be correspondingly larger.  The more likely option,
in our view, is that the ground state near CN is an unpolarized state as predicted by RPA.  Part of the motivation
for this view is the absence of finite-temperature anomalies in experiment, which would signal a phase 
transitions to a paramagnetic state --- expected to be at least weakly first order in MATBG as in other 
itinerant electron magnets \cite{brando2016metallic}.
If so, there is no hint experimentally of the emergence between $\nu=0$ and $|\nu|=1$ 
of the self-consistent Hartree multi-pocket Fermi surface topology illustrated in Fig.~\ref{Fig:SCH_FS}.
Future work should explain why this pocket does not appear (or alternately why its appearance does not influence 
transport), perhaps due to a refinement of the single-particle model which changes flat band wavefunctions \cite{Fang_TBGstrain_2019, Carr_TBGstrain_2019, vafek2023continuum, JKang_TBGcontinuum_2023}, 
exchange interactions within the doped flat bands that stabilize $\kappa,\kappa'$ centered surfaces, 
broken $C_2T$ symmetry related to chiral model physics \cite{tarnopolsky2019origin, khalaf2020soft, khalaf2021charged, bultinck2020ground} and intervalley exchange interactions that we have neglected \cite{bultinck2020ground, Khalaf_4e_2022, Kashiwagi_2022}.  Systematic studies of the evolution of MATBG properties with gate 
induced interlayer displacement fields could play a role in sorting this confusing landscape.


{\em Acknowledgements---} 
Work in Austin was supported by the U.S. Department of Energy, Office of Science, Basic Energy Sciences, under Award DE-SC0019481. M.P. is supported by the European Union's Horizon 2020 research and innovation 
program under the grant agreement No.~881603 - GrapheneCore3 
{and the Marie Sklodowska-Curie grant agreement No.~873028} and by the MUR - Italian Minister of University and Research under 
the ``Research projects of relevant national interest  - PRIN 2020''  - 
Project No.~2020JLZ52N, 
title ``Light-matter interactions and the collective behavior of quantum 2D materials (q-LIMA)''. 
The authors acknowledge resources provided by the Texas Advanced Computing Center (TACC) at The University of Texas at Austin that have contributed to the research results reported in this paper.




%

\clearpage
\onecolumngrid
\widetext
\begin{center}
\textbf{\large Supplemental Material:\\ Weak Coupling Theory of Competing Phases in Magic-Angle Twisted Bilayer Graphene}
\end{center}

\section{Bistritzer-MacDonald model with non-local interlayer tunneling}\label{SM:BM_model}
\begin{figure*}[!h]
\centering
  \includegraphics[scale=1.0]{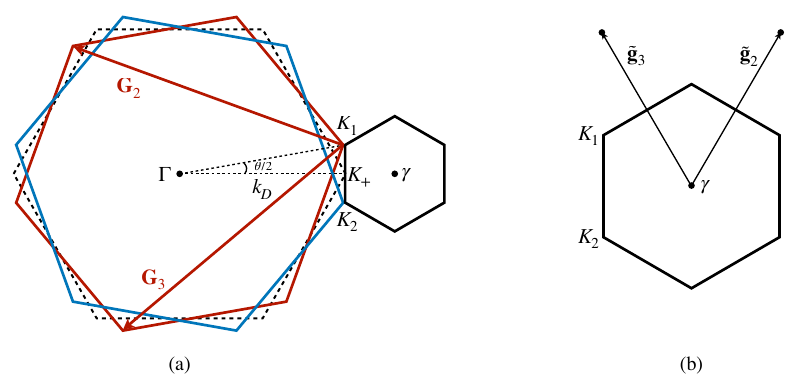}
  \vspace{-5pt}
  \caption{\label{fig_BZs} {\small
  (a) Rotated Brillouin zones of top (red) and bottom (blue) graphene layers. The top layer is rotated anticlockwise with respect to the bottom layer by $\theta$. The black dashed hexagonal is the Brillouin zone of unrotated graphene. $K_+$, $K_1$ and $K_2$ are Dirac points of unrotated graphene, the top graphene layer and the bottom graphene layer respectively. $k_D$ is the length of $\Gamma-K_+$. The smaller black hexagon on the right is the moir\'e Brillouin zone.
  (b) The moir\'e Brillouin zone. Three momentum boosts $\tilde{\mb{g}}_j$ in Eq.~(\ref{Eq_momentum_boosts}) are shown.
  }}
\end{figure*}

In the original Bistritzer-MacDonald (BM) model, Eq. (5) in Ref. \cite{BM}, the interlayer tunneling matrix element is
\be\label{Eq_T}
\begin{split}
T^{\alpha \beta}_{\mb{k} \mb{k}'}
&= \frac{1}{A_{uc}} \sum\limits_{j=1}^3 t_{\mb{k}+\mb{G}_j} e^{i[\mb{G}_j \cdot (\bs{\tau}_\alpha - \bs{\tau}_\beta + \bs{\tau}_0) - \mb{G}_j' \cdot \mb{d}]}
\delta_{\mb{k}+\mb{G}_j, \mb{k}' + \mb{G}_j'} \\
&= \frac{1}{A_{uc}} \sum\limits_{j=1}^3 t_{\mb{k}+\mb{G}_j} T_j^{\alpha \beta}
\delta_{\mb{k}+\mb{G}_j, \mb{k}' + \mb{G}_j'},
\end{split}
\ee
where $\mb{G}$ and $\mb{G}'$ are reciprocal lattice vectors of the top (layer 1) and bottom (layer 2) graphene layers respectively, momenta $\mb{k}$ and $\mb{k}'$ are measured relative to the $\Gamma$ point of the original monolayer graphene and are both near the Dirac point $\mb{K}$. Greek indices label sublattices. The interlayer tunneling $t$ and matrices $T_j$ are defined later.
The top layer (layer 1) is anticlockwise rotated by $\theta/2$ and characterized by vectors without $'$ in the notation, bottom layer (layer 2) is clockwise rotated by $\theta/2$ and characterized by vector with $'$. The reciprocal lattice vectors are related by the rotation operator $\mb{G}'_i = \mathcal{R}_{-\theta} \mb{G}_i$, with
\be
\mathcal{R}_\theta = 
\begin{pmatrix}
\cos \theta & -\sin \theta \\
\sin \theta & \cos \theta
\end{pmatrix}.
\ee
Three $\mb{G}$'s that are most relevant to the interlayer hopping in the two-center approximation are, as shown in Fig.~\ref{fig_BZs}(a),
\be\label{Eq_threeGs}
\begin{split}
&\mb{G}_1 = 
\begin{pmatrix}
0\\
0
\end{pmatrix}, \quad
\mb{G}_2 = \mathcal{R}_{\theta/2} \cdot \frac{4\pi}{\sqrt{3}a}
\begin{pmatrix}
-\frac{\sqrt{3}}{2}\\
\frac{1}{2}
\end{pmatrix}, \quad
\mb{G}_3 = \mathcal{R}_{\theta/2} \cdot  \frac{4\pi}{\sqrt{3}a}
\begin{pmatrix}
-\frac{\sqrt{3}}{2}\\
-\frac{1}{2}
\end{pmatrix}.
\end{split}
\ee
$a = 0.246$ nm is graphene's lattice constant. Starting from AB-stacked bilayer graphene, we choose
\be
\begin{split}
\tau_A = 
\begin{pmatrix}
0\\
0
\end{pmatrix}, \quad
\tau_B = \tau_0 = \mathcal{R}_{\theta/2} \cdot \frac{a}{\sqrt{3}}
\begin{pmatrix}
0\\
1
\end{pmatrix}.
\end{split}
\ee
Three largest interlayer hopping terms, corresponding to three $\mb{G}$'s in Eq.~(\ref{Eq_threeGs}), are
\be
\begin{split}
& T_1 = 
\bmat
u & 1\\
1 & u
\emat, \ \ 
T_2 = e^{-i\mb{G}_2' \cdot \mb{d}}
\bmat
u e^{i\phi} & 1\\
e^{-i\phi} & u e^{i\phi}
\emat, \ \ 
T_3 = e^{-i\mb{G}_3' \cdot \mb{d}}
\bmat
u e^{-i\phi} & 1\\
e^{i\phi} & u e^{-i\phi}
\emat,
\end{split}
\ee
where $\phi = 2\pi/3$, $u=w_{AA}/w_{AB}$ is the ratio of interlayer tunneling between the same sublattice and between different sublattices. The phase factor dependent on $\mb{d}$ can be eliminated by a unitary transformation of the plane-wave expanded basis.

The Fourier transform of interlayer tunneling strength $t(\mb{r})$ in Eq.~(\ref{Eq_T}) is
\be
t_\mb{q} = \int d^2\mb{r} e^{i\mb{q} \cdot \mb{r}} t(\mb{r}).
\ee
Because $t(\mb{r})$ is a smooth and slowly varying function of the projected 2D coordinate $\mb{r}$, $t_\mb{q}$ rapidly decays with respect to $q$ and we only keep three terms in the interlayer tunneling Eq.~(\ref{Eq_T}). Because of the symmetry of carbon $p_z$ orbitals, $t(\mb{r})$ and therefore $t_\mb{q}$ are orientation-independent, i.e. $t_\mb{q} = t_q$.

As shown in Fig.~\ref{fig_BZs}(b), the three allowed momentum boosts, defined by the difference between the momenta in the top and bottom layers $\tilde{\mb{g}}_j = \mb{k} - \mb{k}' = \mb{G}_j' - \mb{G}_j$, are
\be
\label{Eq_momentum_boosts}
\begin{split}
&\tilde{\mb{g}}_1=\begin{pmatrix}
0\\
0
\end{pmatrix}, \quad
\tilde{\mb{g}}_2= \frac{4\pi}{\sqrt{3}a_M}
\begin{pmatrix}
\frac{1}{2}\\
\frac{\sqrt{3}}{2}
\end{pmatrix}, \quad
\tilde{\mb{g}}_3= \frac{4\pi}{\sqrt{3}a_M}
\begin{pmatrix}
-\frac{1}{2}\\
\frac{\sqrt{3}}{2}
\end{pmatrix}, \\
\end{split}
\ee
where $a_M = a/(2 \sin(\theta/2) \approx a/\theta$ is the moir\'e lattice constant.

Momenta $\mb{k}+\mb{G}_j$ in Eq.~(\ref{Eq_T}) are near Brillouin zone corners. To zeroth order,
\be
t_{\mb{k}+\mb{G}_j} \approx t_{\mb{k}_{D,j}} = t_{k_D} = w_0 A_{uc},
\ee
where $\mb{k}_{D,j}$ are Dirac points of the unrotated graphene
\be
\mb{k}_{D,j} = e^{i2\pi(j-1)/3} (1,0) \frac{4\pi}{3a}.
\ee
The interlayer tunneling is local and therefore momentum independent
\be
T_{\mb{k} \mb{k}'} = w_0 \sum\limits_{j=1}^3 T_j \delta_{\mb{k}-\mb{k}', \tilde{\mb{g}}_j}.
\ee 

The approximate particle-hole symmetry is broken by taking into account non-local interlayer tunneling effects\cite{xie2020weak}. Keeping the expansion in $t_{\mb{k}+\mb{G}_j}$ till the first order,
\be
t_\mb{k} \approx t_{k_D} +  t'  (k-k_D),
\ee
where
\be
t' = \frac{dt}{dk} \Big|_{k=k_D} < 0
\ee
is the tunable non-local tunneling parameter.
The momentum-dependent interlayer tunneling becomes
\be
\begin{split}
T_{\mb{k} \mb{k}'} &= \frac{1}{A_{uc}} \sum\limits_{j=1}^3 t_{\mb{k}+\mb{G}_j} T_j \delta_{\mb{k}-\mb{k}',\tilde{\mb{g}}_j}\\
&= \sum\limits_{j=1}^3 \Big[ w_0 + \frac{w_{nl}}{g_M} (|\mb{k}+\mb{G}_j|-k_D)\Big] T_j \delta_{\mb{k}-\mb{k}',\tilde{\mb{g}}_j}.
\end{split}
\ee
The non-local tunneling parameter $w_{nl}$ is defined as
\be
w_{nl} = \frac{g_M}{A_{uc}} t',
\ee
where $g_M=4\pi/\sqrt{3}a_M$ is the length of moir\'e primitive reciprocal lattice vector.

In the numerical calculations in this Letter, we take $v_{_F} = 0.866 \times 10^6$ m/s, $u=w_{AA}/w_{AB}=0.6$, $w_0=110$ meV and $w_{nl} = -20$ meV for MATBG of twist angle $1.1^\circ$. For other efforts to improve the accuracy of the single-particle Hamiltonian of twisted bilayer graphene see Refs.~\cite{Carr_TBGstrain_2019, Fang_TBGstrain_2019, vafek2023continuum, JKang_TBGcontinuum_2023}.

\section{TBG Hamiltonian}\label{SM:full_TBG_Hamil}
The Hamiltonian of a periodic crystal is
\be\label{Eq_Hamil_periodic}
\hat{H} = \hat{T}_e + \hat{H}_{e-e} + \hat{H}_{e-b} + \hat{H}_{b-b},
\ee
where $\hat{T}_e$ is the kinetic energy operator. Instead of the uniform background charge in the Jellium model, a periodic background of the positive charge
\be
n_b(\mb{r}) = \frac{1}{A} \sum\limits_{\mb{G}} n_{b,\mb{G}} e^{i\mb{G} \cdot \mb{r}}
\ee
is assumed. At CN, the background charge cancels with excess electron charge
\be
 n_{b,\mb{0}} = \int n_b(\mb{r}) d\mb{r} = N.
\ee
Written in electron number density operators $\hat{n}({\mb{q}})$ and the total number operator $\hat{N}$
\be
\begin{split}
&\hat{n}({\mb{q}}) = \sum\limits_{\mb{k},\mb{G}} c^\dagger_{\mb{k}+\mb{G}-\mb{q}} c_{\mb{k}+\mb{G}}, \\
&\hat{N} = \hat{n}(\mb{q}=0) = \sum\limits_{\mb{k},\mb{G}} c^\dagger_{\mb{k}+\mb{G}} c_{\mb{k}+\mb{G}},
\end{split}
\ee
the interacting Hamiltonians are
\be
\begin{split}
&\hat{H}_{e-b}
= -e^2 \int d\mb{r} d\mb{r}' \frac{\hat{n}(\mb{r}) n_b(\mb{r}')}{|\mb{r} - \mb{r}'|}
= -\frac{1}{A} \sum\limits_{\mb{G}} V(\mb{G}) n_{b,\mb{G}} \hat{n}_{-\mb{G}}, \\
&\hat{H}_{b-b}
= \frac{e^2}{2} \int d\mb{r} d\mb{r}' \frac{n_b(\mb{r}) n_b(\mb{r}')}{|\mb{r} - \mb{r}'|}
= \frac{1}{2A} \sum\limits_\mb{G} V(\mb{G}) n_{b,\mb{G}} n_{b,-\mb{G}}, \\
& \hat{H}_{e-e}
= \frac{e^2}{2} \sum\limits_{i \neq j} \frac{1}{|\mb{r}_i - \mb{r}_j|}
= \frac{1}{2A} \sum\limits_{\mb{q},\mb{G}} V(\mb{q}+\mb{G}) \big[ \hat{n}({\mb{q}+\mb{G}}) \hat{n}({\mb{-q}-\mb{G}}) - \hat{N}\big].
\end{split}
\ee
Similar to the Jellium model, the $\mb{G}=0$ terms of $\hat{H}_{e-b}$ and $\hat{H}_{b-b}$, and the $\mb{q}=\mb{G}=0$ term of $\hat{H}_{e-e}$ cancel in the thermodynamic limit $A, N \rightarrow \infty$. The full Hamiltonian in Eq.~(\ref{Eq_Hamil_periodic}) can be written as
\be\label{Eq_Hamilee_periodic1}
\begin{split}
\hat{H} 
&= \hat{T}_e + \hat{H}_{e-e} + \hat{V}_b \\
&=\hat{T}_e
+ \frac{1}{2A} \sum\limits'_{\mb{q}, \mb{G}} V(\mb{q}+\mb{G}) \big[ \hat{n}({\mb{q}+\mb{G}}) \hat{n}({-\mb{q}-\mb{G}}) - \hat{N} \big]  + \frac{1}{2A} \sum\limits_{\mb{G} \neq 0}
V(\mb{G}) \Big[-2n_{b,\mb{G}} \hat{n}_{-\mb{G}} + n_{b,\mb{G}}n_{b,-\mb{G}} \Big],
\end{split}
\ee
where $'$ on the summation symbol means $\mb{q}=\mb{G}=0$ term is excluded in the momentum summation.

It is straightforward to generalize the electron-electron interacting Hamiltonian in Eq.~(\ref{Eq_Hamilee_periodic1}) to the TBG case, where two layers and two sublattices degrees of freedom are explicitly included,
\be
\begin{split}
\hat{H}_{e-e}
&= \frac{1}{2A} \sum\limits'_{\substack{\mb{q},\mb{g}, \alpha,\beta \\ \mb{k}_1,\mb{k}_2,\mb{g}_1,\mb{g}_2}} V_{\alpha \beta}(\mb{q}+\mb{g}) c^\dagger_{\alpha,\mb{k}_1+\mb{g}_1-\mb{q}-\mb{g}} 
c^\dagger_{\beta,\mb{k}_2+\mb{g}_2+\mb{q}+\mb{g}} 
c_{\beta,\mb{k}_2+\mb{g}_2} 
c_{\alpha,\mb{k}_1+\mb{g}_1},
\end{split}
\ee
where $\mb{g},\mb{g}_1,\mb{g}_2$ are moir\'e reciprocal lattice vectors and $\mb{k}_1,\mb{k}_2,\mb{q}$ are momenta in the first moir\'e Brillouin zone. $\alpha,\beta$ label layers and sublattices. The $'$ on top of the summation means the $\mb{q}=\mb{g}=0$ term is excluded, which is cancelled by the periodic background of positive charge in the thermodynamic limit.

Alternatively, $\hat{H}_{e-e}$ can be expressed in electron number density operators
\be\label{Eq_Hee_TBG}
\begin{split}
\hat{H}_{e-e}
&=  -\frac{1}{2A}
\sum\limits'_{\mb{q},\mb{g}}
V_{S}(\mb{q}+\mb{g}) \hat{N}
+ \frac{1}{2A} 
\sum\limits'_{\substack{\mb{q},\mb{g} \\ \alpha, \beta}}
V_{\alpha \beta}(\mb{q}+\mb{g})
\sum\limits_{\mb{g}_1,\mb{g}_2}
\hat{n}^{\mb{g}_1}_\alpha(\mb{q}+\mb{g}) 
\hat{n}^{\mb{g}_2}_\beta(-\mb{q}-\mb{g}),
\end{split}
\ee
where
\be
\begin{split}
\hat{N} &= \sum\limits_{\mb{k}, \mb{g},\alpha} c^\dagger_{\alpha, \mb{k}+\mb{g}}
c_{\alpha, \mb{k}+\mb{g}}, \\
\hat{n}(\mb{q}) &= \sum\limits_{\alpha,\mb{g}} 
\hat{n}_\alpha^{\mb{g}}(\mb{q})
= \sum\limits_{\alpha,\mb{g}}
\sum\limits_\mb{k} c^\dagger_{\alpha,\mb{k}+\mb{g}-\mb{q}} c_{\alpha,\mb{k}+\mb{g}}.
\end{split}
\ee
For isotropic dielectric media, the Coulomb interaction within the same 2D layer ($V_S$) and between different 2D layers ($V_D$) are respectively
\be
\label{Eq_VsVd}
\begin{split}
&V_S(\mb{q}) = \frac{2\pi e^2}{\ep q},\\
&V_D(\mb{q}) = \frac{2\pi e^2}{\ep q} e^{-qd},
\end{split}
\ee
where $\ep$ is the relative dielectric constant of the surrounding environment and $d$ is the distance between adjacent layers. 

For anisotropic dielectric media, for example hBN, $\ep_{_{\rm BN}}=\sqrt{\ep_{zz} \ep_\perp} \approx 4.5$ with $\ep_{\perp}=6.9$\cite{hBN_ep_zz}, $\ep_{zz}=3$. 
\be\label{Eq:V_S_V_D}
\begin{split}
&V_S(\mb{q}) = \frac{2\pi e^2}{q\sqrt{\ep_{\perp} \ep_{zz}}},\\
&V_D(\mb{q}) = \frac{2\pi e^2}{q\sqrt{\ep_{\perp} \ep_{zz}}} e^{-qd\sqrt{\ep_{\perp}/\ep_{zz}}}.
\end{split}
\ee
With dual metallic gates that is equidistant from the target 2D system, with distance $d_m$, the screened Coulomb potential energy is
\be
V_{sc}(\mb{q}) 
= \frac{2\pi e^2}{q \sqrt{\ep_\perp \ep_{zz}}} \tanh \Big(qd_m \sqrt{\frac{\ep_\perp}{\ep_{zz}}}\Big).
\ee
In calculations throughout this paper, the $d$-dependence of the Coulomb potential in Eq.~(\ref{Eq:V_S_V_D}) is ignored, {\it{i.e.}},
\be
V(\mb{q}) = V_S(\mb{q}) = V_D(\mb{q}) = \frac{2\pi e^2}{q\ep_{_{\rm BN}}},
\ee 
and the hBN dielectric constant is chosen to be $\epsilon_{_{\rm BN}}=5.1$.
 Then Eq.~(\ref{Eq_Hee_TBG}) simplifies to
\be
\begin{split}
\hat{H}_{e-e} 
&= \frac{1}{2A} \sum\limits'_{\mb{q},\mb{g}} V(\mb{q}+\mb{g}) \Big[ \hat{n}(\mb{q}+\mb{g}) 
\hat{n}(-\mb{q}-\mb{g}) 
- \hat{N} \Big].
\end{split}
\ee

\section{Self-consistent Hartree approximation in TBG}\label{SM:scHartree}
With approximate SU(4) symmetry, the exchange interaction is only between electrons within the same flavor. The inter-flavor interaction is only through the Hartree potential.
The matrix element of the flavor- and momentum-independent Hartree self-energy is
\be
\label{Eq_SigmaH}
\Sigma^H_{\alpha,\mb{g}_1;\beta,\mb{g}_2}
= \frac{\delta_{\alpha \beta}}{A} \sum\limits_{\alpha', \mb{g}_1', \mb{g}_2'} V_{\alpha \alpha'}(\mb{g}_2-\mb{g}_1) \delta \rho_{\alpha',\mb{g}_1';\alpha',\mb{g}_2'} \delta_{\mb{g}_2'-\mb{g}_1',\mb{g}_2-\mb{g}_1},
\ee
where $\delta \bs{\rho}$ is the relative density matrix defined by subtracting the density matrix of the decoupled graphene bilayer at CN
\be
\delta \bs{\rho} = \bs{\rho} - \bs{\rho}_0.
\ee

The flavor-polarization-dependence of the self-energy is implicitly incorporated in the density matrix through the Fermi-Dirac distribution $\Theta(\mu_f - \vep^f_{n\mb{k}})$:
\be
\rho_{\alpha,\mb{g}_1;\beta,\mb{g}_2} = \sum\limits_{f,n,\mb{k}} z^{nf}_{\alpha,\mb{g}_1}(\mb{k}) \bar{z}^{nf}_{\beta,\mb{g}_2}(\mb{k}) \Theta(\mu_f - \vep^f_{n\mb{k}}).
\ee
$f=1,2,3,4$ represent four flavors.
If the inversion symmetry is retained, the diagonal elements of $\Sigma^H$ are constant and can be set to zero.

For any specific flavor polarization $(\nu_1,\nu_2,\nu_3,\nu_4)$, the Hamiltonian is solved self-consistently for each flavor
\be\label{Eq_fullH}
H^f(\mb{k},\nu_1,\nu_2,\nu_3,\nu_4) = H^f_0(\mb{k}) + \Sigma^H(\nu_1,\nu_2,\nu_3,\nu_4),
\ee
where $H^f_0(\mb{k})$ is the single-particle Hamiltonian of flavor $f$.

\begin{figure}
  \centering
  \begin{tabular}{@{}c@{}}    \includegraphics[width=0.8\linewidth]{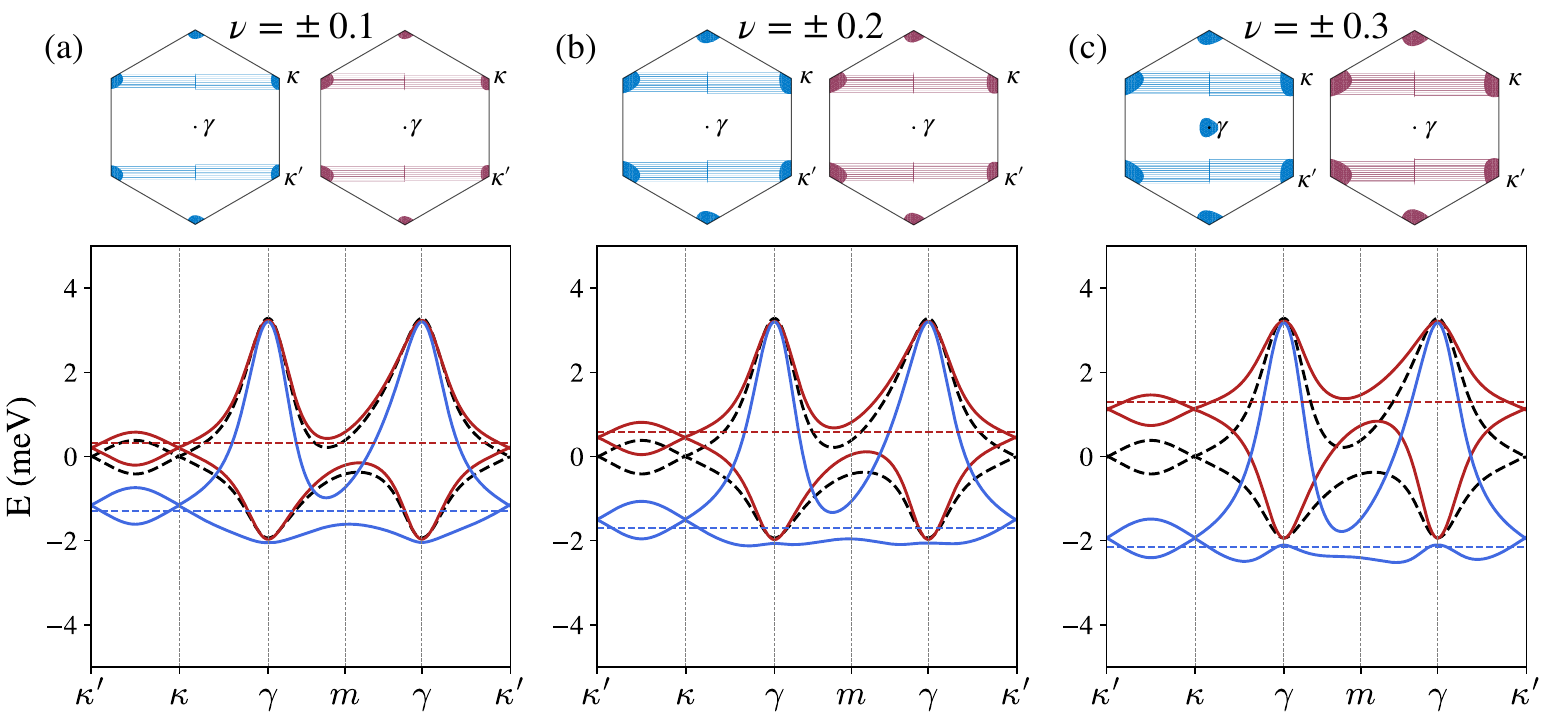} \\[\abovecaptionskip]
  \end{tabular}

  \begin{tabular}{@{}c@{}}    \includegraphics[width=0.8\linewidth]{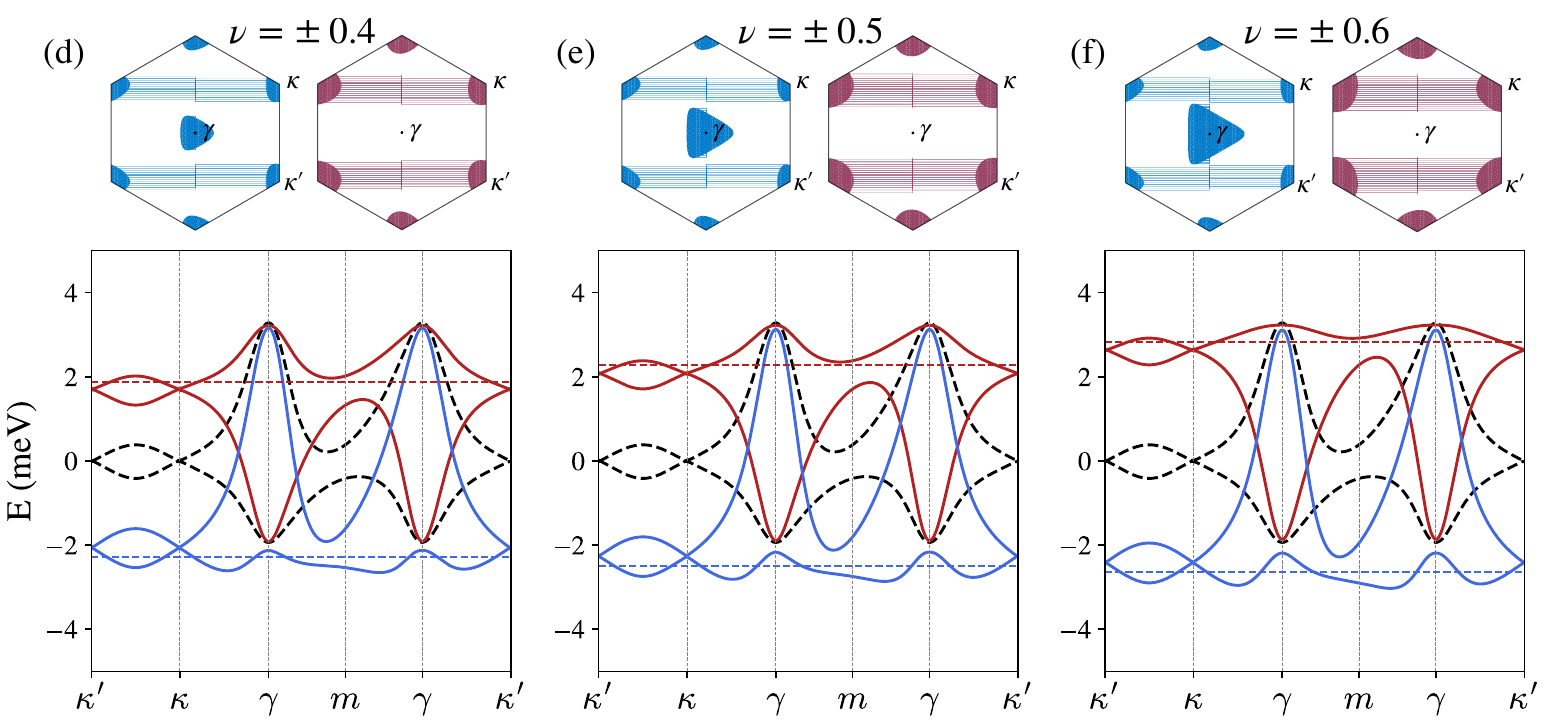} \\[\abovecaptionskip]
  \end{tabular}

  \begin{tabular}{@{}c@{}}    \includegraphics[width=0.8\linewidth]{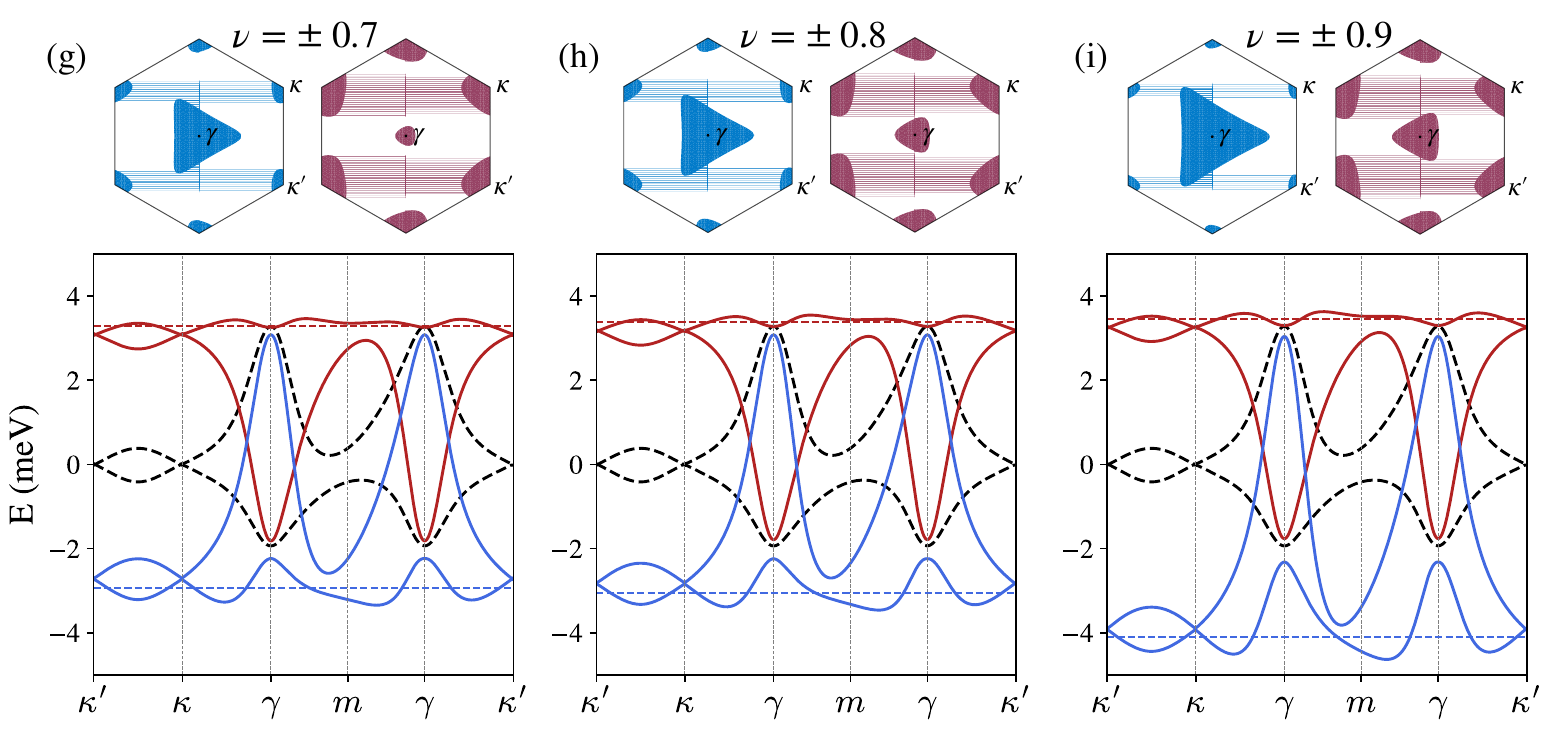} \\[\abovecaptionskip]
  \end{tabular}

  \caption{Self-consistent Hartree band structures (colored lines) and corresponding Fermi surfaces (colored shaded areas) of flavor paramagnetic states at filling factors $\nu \in (-1,1)$. Red (blue) represents electron (hole) doping. The black dashed line in each spectrum is the single-particle band structure. Note that the energy bands and the Fermi levels are shifted such that the non-interacting band energy at $\kappa$ is zero.}\label{FigSM_FS}
\end{figure}

Self-consistent Hartree band structures and corresponding Fermi surfaces of flavor-symmetry unbroken states at filling factors $\nu \in (-1,1)$ are shown in Fig.~\ref{FigSM_FS}.

The electrostatic Hartree energy is
\be
\begin{split}
E_H &= \frac{1}{2} \sum\limits_{f,n,\mb{k}} \langle \Psi_{fn\mb{k}} | \Sigma^H | \Psi_{fn\mb{k}} \rangle \Theta(\mu_f - \varepsilon^f_{n\mb{k}}) \\
&= \frac{1}{2A} \sum\limits_{\substack{\mb{g}_1,\mb{g}_2 \\ \mb{g}'_1,\mb{g}'_2}}
\sum\limits_{\alpha, \alpha'} V_{\alpha \alpha'}(\mb{g}_2-\mb{g}_1) \bar{\rho}_{\alpha,\mb{g}_1;\alpha,\mb{g}_2} \delta \rho_{\alpha',\mb{g}'_1;\alpha',\mb{g}'_2} \delta_{\mb{g}_2'-\mb{g}_1',\mb{g}_2-\mb{g}_1},
\end{split}
\ee
and is further regularized by subtracting the negative Fermi sea contribution
\be
\label{Eq_EH}
E_H = \frac{1}{2A} \sum\limits_{\substack{\mb{g}_1,\mb{g}_2 \\ \mb{g}'_1,\mb{g}'_2}}
\sum\limits_{\alpha, \alpha'} V_{\alpha \alpha'}(\mb{g}_2-\mb{g}_1) \delta \bar{\rho}_{\alpha,\mb{g}_1;\alpha,\mb{g}_2} \delta \rho_{\alpha',\mb{g}'_1;\alpha',\mb{g}'_2} \delta_{\mb{g}_2'-\mb{g}_1',\mb{g}_2-\mb{g}_1}.
\ee
The SCH energy $E_0$ in the main text is defined to include both the SCH band dispersion and the electrostatic Hartree energy:
\be
\label{Eq_E0}
\begin{split}
E_0 &= E_{\text{band}} - E_H \\
&= \sum\limits_{f,n,\mb{k}} \varepsilon^f_{n\mb{k}} \Theta(\mu_f - \varepsilon^f_{n\mb{k}}) - E_H,
\end{split}
\ee
where $\varepsilon^f_{n\mb{k}}$ are eigenvalues of SCH Hamiltonian Eq.~(\ref{Eq_fullH}) and $E_H$ is subtracted to avoid double-counting of the Coulomb energy.

\section{Self-consistent Hartree-Fock Fermi surfaces}\label{SM:scHF}
\begin{figure}
\includegraphics[width=0.82\linewidth]{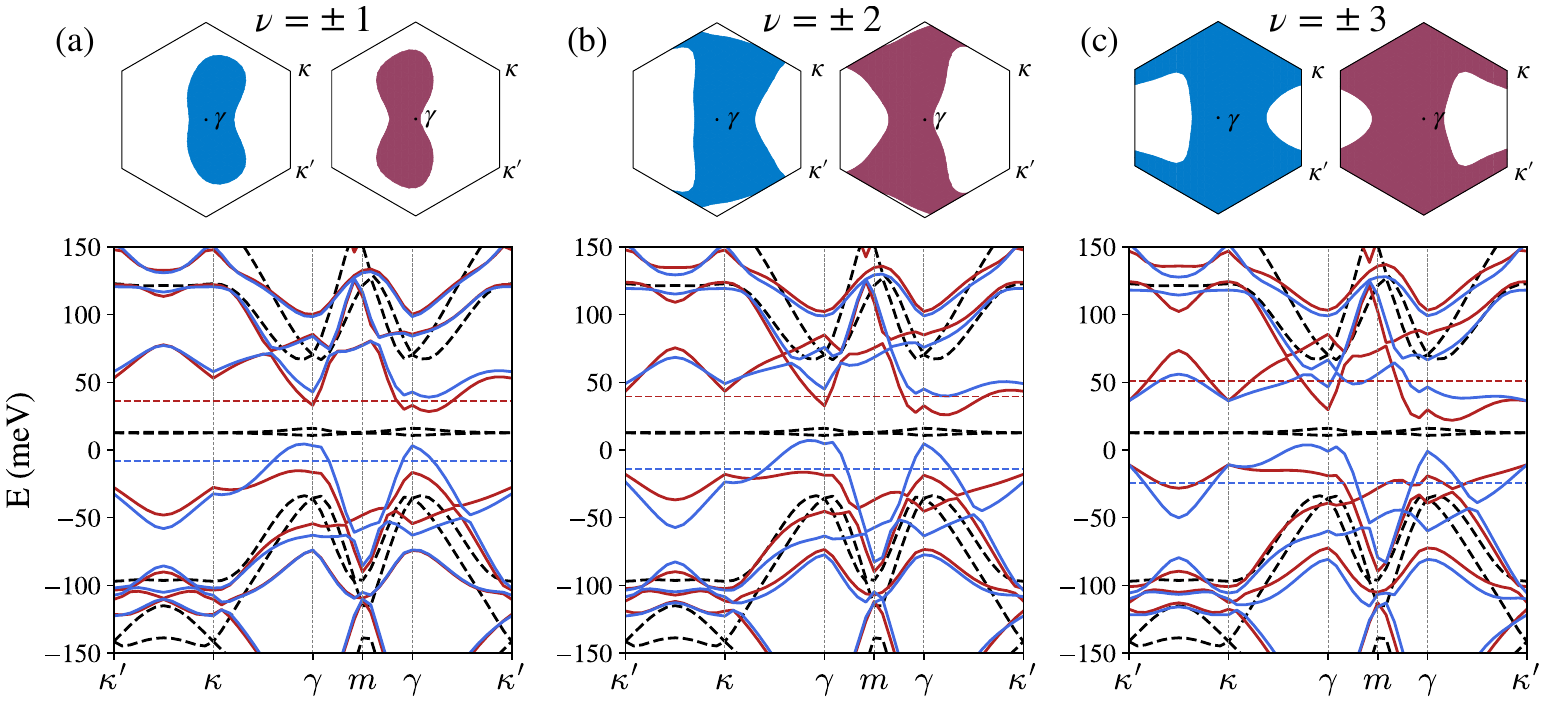}
  \caption{\small Self-consistent Hartree-Fock band structures (colored lines) and corresponding Fermi surfaces (colored shaded areas) of flavor paramagnetic states at filling factors $\nu = \pm 1, \pm 2, \pm 3$. Red (blue) represents electron (hole) doping. The black dashed line in each spectrum is the single-particle band structure.}\label{FigSM_FS_HF}
\end{figure}

To compare with the self-consistent Hartree approximation, we present in Fig.~\ref{FigSM_FS_HF} the Fermi surfaces of self-consistent Hartree-Fock calculations with momentum- and flavor-dependent Fock self-energy
\be
\label{Eq_SigmaH}
\Sigma^{F,f}_{\alpha,\mb{g}_1;\beta,\mb{g}_2} (\mb{k})
= -\frac{1}{A} \sum\limits_{\mb{k}', \mb{g}_1', \mb{g}_2'} V_{\alpha \beta}(\mb{k}'-\mb{k}+\mb{g}'_1-\mb{g}_1) \delta \rho^f_{\alpha,\mb{g}_1';\beta,\mb{g}_2'}(\mb{k}') \delta_{\mb{g}_2'-\mb{g}_1',\mb{g}_2-\mb{g}_1}.
\ee

\section{The density response function of TBG}\label{SM:chi_symmetry}
At zero temperature, the matrix elements of the flavor-specified density response function $\tilde{\chi}^f$ of MATBG can be derived following the Lindhard formula
\be
\label{Eq_chi_T0}
\begin{split}
[\tilde{\chi}^f]^{\mb{g} \mb{g}'}(\mb{q},\omega) 
= \frac{1}{A} \sum\limits_{n,m,\mb{k}} \frac{\Theta^f_{n\mb{k}} - \Theta^f_{m\mb{k}+\mb{q}}}{\omega + \varepsilon^f_{n\mb{k}} - \varepsilon^f_{m\mb{k}+\mb{q}} + i \eta}
\Big[ \sum\limits_{\alpha,\mb{g}_1} z^{nf}_{\alpha,\mb{g}_1}(\mb{k})
\bar{z}^{mf}_{\alpha,\mb{g}_1+\mb{g}}(\mb{k}+\mb{q}) \Big]^*
\sum\limits_{\beta,\mb{g}_2}
z^{nf}_{\beta,\mb{g}_2}(\mb{k})
\bar{z}^{mf}_{\beta,\mb{g}_2+\mb{g}'}(\mb{k}+\mb{q}).
\end{split}
\ee
where $\varepsilon$ and $z$ are quasiparticle eigen-energy and eigenvector of the self-consistent Hartree approximation.
It can be easily proved that Eq.~(\ref{Eq_chi_T0}) satisfies the symmetry
\be
\label{Eq_chi_sym_re}
[\tilde{\chi}^f]^{\mb{g} \mb{g}'}(\mb{q},\omega) = \big[[\tilde{\chi}^f]^{-\mb{g} -\mb{g}'}(-\mb{q},-\omega)\big]^*.
\ee
It obeys the general property of the response function
\be
\tilde{\chi}_{AB}(\omega) = [\tilde{\chi}_{A^\dagger B^\dagger}(-\omega)]^*.
\ee
The density response functions of opposite valleys with the same electrostatic doping level are related by the spinless time-reversal symmetry
\be
[\tilde{\chi}^-]^{\mb{g} \mb{g}'}(\mb{q},\omega) = [\tilde{\chi}^+]^{-\mb{g}' -\mb{g}}(-\mb{q},\omega),
\ee
following reciprocity relations
\be
\tilde{\chi}^-_{AB}(\omega) = \tilde{\chi}^+_{B^TA^T}(\omega),
\ee
where $\pm$ represent two opposite valleys in TBG.
If the TBG system is flavor unpolarized, the total density response function $\tilde{\chi}$ summed over four flavors is time-reversal invariant and satisfies
\be
\begin{split}
\tilde{\chi}^{\mb{g} \mb{g}'}(\mb{q},\omega) 
&= 2 [\tilde{\chi}^+]^{\mb{g} \mb{g}'}(\mb{q},\omega) + 2 [\tilde{\chi}^-]^{\mb{g} \mb{g}'}(\mb{q},\omega) \\
&= 2 [\tilde{\chi}^+]^{\mb{g} \mb{g}'}(\mb{q},\omega) + 2 [\tilde{\chi}^+]^{-\mb{g}' -\mb{g}}(-\mb{q},\omega) \\
&= \tilde{\chi}^{-\mb{g}' -\mb{g}}(-\mb{q},\omega).
\end{split}
\ee
At finite temperatures, the density response function is
\be\label{Eq:chiH}
\begin{split}
[\tilde{\chi}^f]^{\mb{g} \mb{g}'}(\mb{q},i\omega) 
= \frac{1}{A} \sum\limits_{n,m,\mb{k}} &\frac{\Theta^f_{n\mb{k}}-\Theta^f_{m\mb{k}+\mb{q}}}{i\omega + \varepsilon^f_{n\mb{k}} - \varepsilon^f_{m\mb{k}+\mb{q}}} 
\Big[ \sum\limits_{\alpha,\mb{g}_1} z^{nf}_{\alpha,\mb{g}_1}(\mb{k})
\bar{z}^{mf}_{\alpha,\mb{g}_1+\mb{g}}(\mb{k}+\mb{q}) \Big]^*
\sum\limits_{\beta,\mb{g}_2} 
z^{nf}_{\beta,\mb{g}_2}(\mb{k})
\bar{z}^{mf}_{\beta,\mb{g}_2+\mb{g}'}(\mb{k}+\mb{q}).
\end{split}
\ee
Matrix $\tilde{\chi}^f(\mb{q},i\omega)$ is Hermitian along the imaginary frequency axis, {\it i.e.},
\be
\begin{gathered}
\tilde{\chi}^f(\mb{q},i\omega) = [\tilde{\chi}^f]^\dagger(\mb{q},-i\omega),\\
[\tilde\chi^f]^{\mb{g} \mb{g}'}(\mb{q},i\omega) = \big[[\tilde\chi^f]^{\mb{g}' \mb{g}}(\mb{q},-i\omega)\big]^*.
\end{gathered}
\ee
$\tilde{\chi}^f(\mb{q},i\omega)$ also satisfies
\be
\label{chi_GGprime_property2}
[\tilde\chi^f]^{\mb{g} \mb{g}'}(\mb{q},i\omega) = [\tilde\chi^f]^{-\mb{g}' -\mb{g}}(-\mb{q},-i\omega) = \big[ [\tilde\chi^f]^{-\mb{g} -\mb{g}'}(-\mb{q},i\omega) \big]^*.
\ee
Opposite valleys with the same electrostatic doping level are related by the time-reversal symmetry
\be
[\tilde{\chi}^-]^{\mb{g}\mb{g}'}(\mb{q},i\omega) 
= [\tilde{\chi}^+]^{-\mb{g}'-\mb{g}}(-\mb{q},i\omega) = \big[[\tilde{\chi}^+]^{\mb{g}'\mb{g}}(\mb{q},i\omega) \big]^* = \big[[\tilde{\chi}^-]^{\mb{g}'\mb{g}}(\mb{q},-i\omega) \big]^*.
\ee
For valley unpolarized state, the total proper density response function in Eq.~(\ref{Eq:chiH4f}) satisfies
\be
\begin{gathered}
\tilde{\chi}^{\mb{g} \mb{g}'}(\mb{q}, i\omega) 
= 2[\tilde{\chi}^+]^{\mb{g} \mb{g}'}(\mb{q},i\omega) + 2[\tilde{\chi}^-]^{\mb{g} \mb{g}'}(\mb{q},i\omega) 
= 2[\tilde{\chi}^+]^{\mb{g} \mb{g}'}(\mb{q},i\omega) + 2\big[ [\tilde{\chi}^+]^{\mb{g}' \mb{g}}(\mb{q},i\omega) \big]^*,
\end{gathered}
\ee
and therefore the density response function of flavor unpolarized state is Hermitian, {\it i.e.},
\be
\tilde{\chi}(\mb{q},i\omega) =\tilde{\chi}^\dagger(\mb{q},i\omega).
\ee
We see that symmetries of $\tilde{\chi}(\mathbf{q}, i\omega)$ are different from those of $\tilde{\chi}(\mathbf{q}, \omega)$ at zero temperature.

\section{The Coupling-constant integration as an evaluation of the ground state energy}\label{SM:coupling_constant_integration}
The ground-state energy of an electron system can be easily connected to the density-density linear-response function---for example at the level of the popular RPA --- by the integration over the coupling constant theorem (see, for example, Sect.~I.8.3 of Ref.~\cite{Giuliani_Vignale}). 
However, the ground-state electron density of a moir\'{e} crystal is inhomogeneous on the moir\'{e} superlattice length scale.
The application of the integration over the coupling constant theorem to an inhomogeneous many-body system is not straightforward and presents some subtleties. On the contrary, the Hohenberg-Kohn and Kohn-Sham theorems~(see, for example, Chapter 7 of Ref.~\cite{Giuliani_Vignale}) of density functional theory are the natural theoretical framework to deal with inhomogeneous many-body electron systems. 
This is why we formulate the problem of the calculation of the ground-state energy of twisted bilayer graphene within the framework of Adiabatic Connection Fluctuation and Dissipation Theorem (ACFDT)~\footnote{See e.g. Ref.[\onlinecite{ACFDT_Gorling_2019}] and references therein to earlier work.}
and then recover the RPA by taking a suitable limit. This theory applies the integration over the coupling constant theorem between the Kohn-Sham (KS) ground state and the real ground state of the electronic Hamiltonian as detailed in the following.

\subsection{The Adiabatic Connection Fluctuation and Dissipation Theorem}
The electronic Hamiltonian can be written as
\begin{equation}\label{electronic_hamiltonian}
\hat{\cal H} = \hat{T}_{\rm e} + \hat{\cal H}_{\rm e-e}+ \hat{V}_{\rm ext}~,
\end{equation}
where $\hat{T}_{\rm e }$ is the kinetic operator, $\hat{\cal H}_{\rm e-e}$ is the electron-electron interaction, i.e.
\begin{equation}\label{electron-electron}
\begin{split}
\hat{H}_{\rm e-e} 
&= \frac{1}{2A} \sum\limits'_{\mb{q},\mb{G}} V(\mb{q}+\mb{G}) \Big[ \hat{n}(\mb{q}+\mb{G}) 
\hat{n}(-\mb{q}-\mb{G}) 
- \hat{N} \Big],
\end{split}
\end{equation}
and $\hat{V}_{\rm ext}$ is the crystal potential, i.e. 
\begin{equation}
\hat{V}_{\rm ext} = \int d^2 \mb{r} V_{\rm ext}(\mb{r})\hat{n}(\mb{r}) ~.
\end{equation}
In Eq.~(\ref{electron-electron}), $V({\mb q +\mb G})$ is the Fourier transform of the electron-electron interaction potential, evaluated at the wave vector ${\mb q +\mb G}$, where ${\mb q}$ is in the first BZ and ${\mb G}$ is an arbitrary reciprocal lattice vector. Similarly, $\hat{n}(\mb q+ \mb G)$ is the Fourier transform of the ground-state density operator $\hat{n}({\mb r})$.

Let $|\Phi\rangle$ be the ground state of the full Hamiltonian and $n(\mb r)$ the associated ground-state density, i.e.
\begin{equation}
n(\mb r) = \langle \Phi|\hat{n}(\mb r)|\Phi\rangle~.
\end{equation}
We now introduce a key auxiliary system, which is described by the so-called {\it Kohn-Sham Hamiltonian} $\hat{\cal H}_{\rm KS}$:
\begin{equation}
\hat{\cal H}_{\rm KS} = \hat{T}_{\rm e } + \hat{V}_{\rm ext} + \hat{V}_{\rm H} + \hat{V}_{\rm xc} \equiv \hat{T}_{\rm e } + \hat{V}_{\rm KS}~.
\end{equation}
This Hamiltonian is a (self-consistent) one-particle Hamiltonian whose ground state $|\psi_{\rm KS}\rangle$ is a Slater determinant of Kohn-Sham orbitals. The fundamental property of this Hamiltonian is that {\it it yields the exact same density of the full Hamiltonian} (\ref{electronic_hamiltonian}).

We now define a family $\hat{\cal H}_\lambda$ of Hamiltonians depending on a real dimensionless parameter $\lambda \in [0,1]$:
\begin{equation}\label{eq:Hlambda}
\hat{\cal H}_\lambda = \hat{T}_{\rm e } + \lambda \hat{\cal H}_{\rm e-e}+ \hat{V}_\lambda~.
\end{equation}  
Let $|\psi_\lambda \rangle$ be the normalized ground state of $\hat{\cal H}_\lambda$, i.e.~$\langle \psi_\lambda |\psi_\lambda \rangle=1~\forall \lambda$.
In Eq.~(\ref{eq:Hlambda}), $\hat{T}_{\rm e}$ and $\hat{\cal H}_{\rm e-e}$ have the exact same meaning as above. The key new quantity is $\hat{V}_\lambda$, which is a local potential that interpolates between the Kohn-Sham potential in the limit $\lambda=0$, i.e.~$\hat{V}_{\lambda=0} =  \hat{V}_{\rm KS}$, and the exact, physical crystal potential in the limit $\lambda =1$, i.e.~$\hat{V}_{\lambda=1} =  \hat{V}_{\rm ext}$. Crucially, $\hat{V}_\lambda$ varies with $\lambda$ in such a way that the correct electronic density is reproduced at every value of $\lambda$, i.e.
\begin{equation}
n_\lambda(\mb r) \equiv \langle\psi_\lambda|\hat{n}(\mb r)|\psi_\lambda \rangle = n(\mb r)~\forall \lambda~.
\end{equation}
The uniqueness (up to a constant) of $\hat{V}_\lambda$ can be proved by applying the Hohenberg-Kohn theorem to the electronic Hamiltonian with reduced coupling constant. In the following, we will fix the arbitrary constant by setting the average value of $\hat{V}_\lambda$ to zero.

The ground state energy at coupling constant $\lambda$ is
\begin{equation}
E(\lambda) =\langle \psi_\lambda |\hat{\cal H}_\lambda| \psi_\lambda\rangle~.
\end{equation}
By applying the Hellman-Feynman theorem we obtain
\begin{equation}
\frac{dE(\lambda)}{d\lambda} =\langle \psi_\lambda |\hat{\cal H}_{\rm e-e}| \psi_\lambda\rangle + \int d^2 \mb r n(\mb r) \partial_\lambda V_\lambda(\mb r)~.
\end{equation}
Integrating the previous differential equation between $\lambda= 0$ and $\lambda=1$ we find:
\begin{equation}\label{eq:integration_over_lambda_1_0}
E(1)-E(0) =\int_0^1 d\lambda \langle \psi_\lambda |\hat{\cal H}_{\rm e-e}| \psi_\lambda\rangle + \int d^2 \mb r n(\mb r) [ V_{\rm ext}(\mb r)-V_{\rm KS}(\mb r)]~.
\end{equation}
In writing the previous two equations we made use of the crucial fact that $n_\lambda(\mb r)=n(\mb r)$ for every $\lambda $ in the integration interval.

We now recall that, in DFT, the exact ground-state energy $E$ of the system, which in the notation of Eq.~(\ref{eq:integration_over_lambda_1_0}) coincides with $E(1)$, is given by~\cite{Giuliani_Vignale}
\begin{equation}\label{eq:E1book}
E = E(1) = T_{\rm s} + \int d^2 \mb r n(\mb r) V_{\rm ext}(\mb r)+ E_{\rm H} + E_{\rm xc}~,
\end{equation}
where $T_{\rm s}$ is the non-interacting kinetic energy functional, i.e.~the kinetic energy of a non-interacting system whose ground-state density is $n({\mb r})$, $E_{\rm H}$ is the Hartree energy, i.e.
\begin{equation}\label{eq:Hartree}
E_{\rm H} = \frac{1}{2}\int d^2 {\mb r}\int d^2 {\mb r}' V(|{\mb r}- {\mb r}'|)n({\mb r})n({\mb r}')~,
\end{equation}
and $E_{\rm xc}$ is the exchange-correlation energy functional.

On the other hand, the quantity $E(0)$ is the average of the Kohn-Sham Hamiltonian over the Kohn-Sham ground-state 
\begin{equation}\label{eq:E0book}
E(0) =\sum_{\alpha \in \rm occ.} \epsilon_\alpha^{\rm KS} = T_{\rm s} + \int d^2 \mb r n(\mb r) V_{\rm KS}(\mb r)~,
\end{equation}
where $\epsilon_\alpha^{\rm KS}$ are the eigenvalues of the Kohn-Sham equations and the sum runs over the occupied states.

Taking the difference between Eq.~(\ref{eq:E1book}) and Eq.~(\ref{eq:E0book}) and comparing the result with Eq.~(\ref{eq:integration_over_lambda_1_0}) we find the following important result:
\begin{equation}
E_{\rm H} + E_{\rm xc} = \int_0^1 d\lambda\langle \psi_\lambda |\hat{\cal H}_{\rm e-e}| \psi_\lambda\rangle~.
\end{equation}

We can conveniently express the matrix element $\langle\psi_\lambda |\hat{\cal H}_{\rm e-e}| \psi_\lambda\rangle$ of the interaction Hamiltonian in terms of the density-density response function by using the fluctuation-dissipation theorem~\cite{Giuliani_Vignale}. At zero temperature and assuming a non-degenerate ground state, we obtain
\begin{equation}\label{fluctuation_dissipation}
\begin{split}
\langle \psi_\lambda |\hat{n}_{\mb q+\mb G}\hat{n}_{-\mb q-\mb G}|\psi_\lambda \rangle & = -\frac{\hbar}{\pi}\int_0^\infty  {\rm Im} \left[\chi_{\hat{n}_{\mb q+\mb G}\hat{n}_{-\mb q-\mb G}}(\omega,\lambda)\right]d\omega +\langle \Phi_\lambda | \hat{n}_{\mb q+\mb G} |\Phi_\lambda \rangle\langle \Phi_\lambda | \hat{n}_{-\mb q-\mb G} |\Phi_\lambda \rangle\\
 &= -\frac{\hbar A}{\pi}\int_0^\infty  {\rm Im}\left[\chi_{nn}^{\mb G\mb G}(\mb q, \omega,\lambda)\right]d\omega +n_{\mb q+\mb G} n_{-\mb q-\mb G}~.
\end{split}
\end{equation}
Making use of (\ref{fluctuation_dissipation}) and (\ref{electron-electron}) we get
\begin{equation}
\langle \psi_\lambda |\hat{\cal H}_{\rm e-e}| \psi_\lambda\rangle = \frac{N}{2}\sum_{\mb G}\int_{\rm BZ} \frac{d^2 \mb q}{(2\pi)^2}V_{\mb q +\mb G}\left\{\frac{n_{\mb q+\mb G}n_{-\mb q -\mb G}}{N}-\frac{\hbar}{n\pi}\int_0^\infty d\omega {\rm Im}[\chi^{\mb G \mb G}_{nn}(\mb q, \omega,\lambda)]-1\right\}.
\end{equation} 
The first term is independent of $\lambda$ and coincides with the Hartree energy,
\begin{equation}
E_{\rm H} = \frac{1}{2A}\sum_{\mb G \neq {\mb 0}}V_{\mb G}n_{\mb G}n_{-\mb G}~.
\end{equation}
We are therefore left with
\begin{equation}
 E_{\rm xc} = \int_0^1 d\lambda \frac{N}{2}\sum_{\mb G}\int \frac{d^2 \mb q}{(2\pi)^2}V_{\mb q +\mb G}\left\{-\frac{\hbar}{n\pi}\int_0^\infty d\omega {\rm Im}[\chi^{\mb G \mb G}_{nn}(\mb q, \omega,\lambda)]-1\right\}~.
\end{equation}
It can be further shown that the exchange energy can be written as
\begin{equation}
E_{\rm x} = \frac{N}{2}\sum_{\mb G}\int \frac{d^2 \mb q}{(2\pi)^2}V_{\mb q +\mb G}\left\{-\frac{\hbar}{n\pi}\int_0^\infty d\omega {\rm Im}[\chi^{\mb G \mb G}_{\rm KS}(\mb q, \omega)]-1\right\}~,
\end{equation}
where $\chi^{\mb G \mb G}_{\rm KS}(\mb q, \omega)$ is the Kohn-Sham response function. Note that this is the exchange energy calculated on the KS orbitals, which is {\it different} from the Hartree-Fock exchange.

The difference between $E_{\rm xc}$ and $E_{\rm x}$ is the correlation energy:
\begin{equation}
 E_{\rm c}= \frac{N}{2}\int_0^1 d\lambda \sum_{\mb G}\int_{\rm BZ} \frac{d^2 \mb q}{(2\pi)^2}V_{\mb q +\mb G}\left\{-\frac{\hbar}{n\pi}\int_0^\infty d\omega {\rm Im}[\chi^{\mb G \mb G}_{nn}(\mb q, \omega,\lambda)-\chi_{\rm KS}^{\mb G \mb G}(\mb q,\omega)]\right\}~.
\end{equation}

In a crystal, the linear response relation---see Eq.~(7.182) in Ref.~\cite{Giuliani_Vignale}---relating the full density response at coupling constant $\lambda$ to the Kohn-Sham density response reads as following:
\begin{equation}\label{eq:linear_response}
\delta n_{\mb q +\mb G}(\omega)=\sum_{\mb G'}\chi_{\rm KS}^{\mb G\mb G'}(\mb q,\omega) \left\{V^{\rm ext}_{\mb q+\mb G'}(\omega)+\sum_{\mb G''} [\lambda V_{\mb q + \mb G'}\delta_{\mb G'\mb G''} + f_{\rm xc, L}^{\mb G' \mb G''}(\mb q,\omega, \lambda)] \delta n_{\mb q +\mb G''}(\omega)\right\},
\end{equation}
where $f_{\rm xc, L}^{\mb G' \mb G''}(\mb q,\omega, \lambda)$ is the wave vector and frequency-dependent exchange-correlation kernel~\cite{Giuliani_Vignale} evaluate at coupling constant $\lambda$. Treating functions in Eq.~(\ref{eq:linear_response}) as matrices with respect to reciprocal lattice vectors indices, we can finally rewrite Eq.~(\ref{eq:linear_response}) as
\begin{equation}\label{eq:linear_response_mat}
\delta n =\chi_{\rm KS} \cdot \left\{V^{\rm ext}+[\lambda V + f_{\rm xc, L}(\lambda)] \cdot \delta n\right\}~.
\end{equation}
Using the same notation, the definition of the full response function at coupling constant $\lambda$ reads as following:
\begin{equation}
\delta n =\chi (\lambda)\cdot V^{\rm ext}~.
\end{equation}
Substituting this definition into Eq.~(\ref{eq:linear_response_mat}) yields 
\begin{equation}
\chi (\lambda) = \chi_{\rm KS}\cdot \left\{1+[\lambda V+f_{\rm xc,L}(\lambda) ]\cdot \chi (\lambda) \right\}~.
\end{equation}
Carrying out some simple algebraic manipulation we finally find
\begin{equation}
\chi (\lambda) = \chi_{\rm KS} \cdot \{1-[\lambda V+f_{\rm xc,L}(\lambda) ]\cdot \chi_{\rm KS}  \}^{-1}~,
\end{equation}
which can be further rearranged into
\begin{equation}
\chi(\lambda)-\chi_{\rm KS} = \chi_{\rm KS} \cdot [\lambda V+f_{\rm xc,L}(\lambda)] \cdot \chi_{\rm KS} \cdot \{1-[\lambda V+f_{\rm xc,L}(\lambda) ]\cdot \chi_{\rm KS}  \}^{-1}~.
\end{equation}
Substituting into the formula for the correlation energy we get the exact expression
\begin{equation}
\label{eq:Ec_full}
E_{\rm c} = -\frac{\hbar N}{2\pi n}\int_{\rm BZ}\frac{d^2\mb q}{(2\pi)^2}\int_0^\infty d\omega \int_0^1 d\lambda {\rm Im}\left\{ \left[V\cdot \chi_{\rm KS} \cdot [\lambda V+f_{\rm xc,L}(\lambda)] \cdot \chi_{\rm KS} \cdot [1-(\lambda V+f_{\rm xc,L}(\lambda)) \cdot \chi_{\rm KS}  ]^{-1}\right](\mb q,\omega) \right\}.
\end{equation}
The total energy is finally given by
\begin{equation}\label{eq:GS_energy}
E = \sum_{\alpha \in \rm occ.} \epsilon_\alpha^{\rm KS} -\int d^2 \mb r n(\mb r) V_{\rm xc}(\mb r) - E_{\rm H}+ E_{\rm x} + E_{\rm c}~.
\end{equation}

\subsection{Taking the RPA limit}
The RPA is obtained by setting to zero $\hat{V}_{\rm xc}$ in the KS equations and $f_{\rm xc,L}$ in (\ref{eq:Ec_full}). In this limit, the KS equations become the Hartree equations, the KS orbitals become the Hartree orbitals, and the KS response function becomes the Hartree response function.

The coupling-constant integral can then be done analytically, yielding
\begin{equation}
\begin{split}
E_{\rm c,RPA} & = -\frac{\hbar N}{2\pi n}\int \frac{d^2\mb q}{(2\pi)^2}\int_0^\infty d\omega \int_0^1 d\lambda \lambda \rm Im \Big\{\text{Tr}\big[ [ V \cdot \chi_{\rm H} ]^2 \cdot [1-\lambda V \cdot \chi_{\rm H}  ]^{-1}\big](\mb q,\omega)\Big\} \\
& =-\frac{\hbar N}{2\pi n}\int \frac{d^2\mb q}{(2\pi)^2}\int_0^\infty d\omega \int_0^1 d\lambda \lambda \rm Im \Big\{ \text{Tr}\big[(\sqrt{V}\cdot \chi_{\rm H}\cdot\sqrt{V})^2\cdot (1-\lambda \sqrt{V}\cdot \chi_{\rm H}\cdot\sqrt{V})^{-1}\big](\mb q,\omega)\Big\} \\
& =\frac{N}{2 n} \int \frac{d^2 \mb q}{(2\pi)^2} \int_0^\infty \frac{\hbar d\omega}{\pi} \text{Im} \Big\{ \text{Tr}[\sqrt{V}\cdot \chi_{\rm H}\cdot\sqrt{V}+\ln(1-\sqrt{V}\cdot \chi_{\rm H}\cdot\sqrt{V})](\mb q,\omega)\Big\}.
\end{split}
\end{equation}
Setting $\hat{V}_{\rm xc}= 0$ in (\ref{eq:GS_energy}) we obtain the final expression for the RPA (or time-dependent Hartree) ground-state energy of an inhomogeneous system,
\begin{equation}
E = \sum_\alpha^{\rm occ.} \epsilon_\alpha^{\rm H} - E_{\rm H} + E_{\rm x} + E_{\rm c, RPA}~.
\end{equation}
Note that the Hartree and exchange energies are now calculated on the Hartree orbitals.
The first two terms in the previous equation coincide with the Hartree expression of the ground-state energy, avoiding double counting of the Coulomb interaction energy.


\section{The exchange-correlation energy of TBG}\label{SM:GWA_Exc}

Following the ACFDT in SM~\ref{SM:coupling_constant_integration},
the exchange-correlation energy of TBG is,
\be
\label{Eq_Exc_TBG2}
\begin{split}
E_{xc} 
&= \frac{n}{2} \sum\limits'_{\mb{q},\mb{g}} V(\mb{q}+\mb{g}) \Big[ -\frac{1}{\pi n} \int_0^1 d\lambda \int_0^\infty \text{Im} \chi^{\mb{g} \mb{g}}(\mb{q},\omega;\lambda) d\omega -1 \Big] \\
&= \frac{n}{2} \sum\limits'_{\mb{q},\mb{g}} V(\mb{q}+\mb{g}) \Big[ -\frac{1}{\pi n} \int_0^1 d\lambda \int_0^\infty \text{Re} \chi^{\mb{g} \mb{g}}(\mb{q},i\omega;\lambda) d\omega -1 \Big] \\
&= \frac{n}{2} \sum\limits'_{\mb{q},\mb{g}} V(\mb{q}+\mb{g}) \Big[ -\frac{1}{\pi n} \int_0^1 d\lambda \int_0^\infty \chi^{\mb{g} \mb{g}}(\mb{q},i\omega;\lambda) d\omega -1 \Big].
\end{split}
\ee
In the last two expressions above, the integral along the real axis is rotated to the imaginary axis using the contour deformation, which is justified below.

The response function can be expressed in the entire complex plane using the spectral representation,
\be
\chi (z) = -\frac{1}{\pi} \int_{-\infty}^\infty \frac{\text{Im} \chi(\omega)}{z-\omega} d\omega.
\ee
In TBG, the density response function satisfies (as in Eq.~(\ref{Eq_chi_sym_re}))
\be
\tilde{\chi}^{\mb{g} \mb{g}'}(\mb{q},\omega) = \big[ \tilde{\chi}^{-\mb{g} -\mb{g}'}(-\mb{q},-\omega) \big]^*,
\ee
therefore
\be
\begin{split}
\tilde{\chi}^{\mb{g} \mb{g}'}(\mb{q},z) 
&= -\frac{1}{\pi} \int_0^\infty \Bigg[ \frac{\text{Im} \tilde{\chi}^{\mb{g} \mb{g}'}(\mb{q},-\omega)}{z+\omega} + \frac{\text{Im} \tilde{\chi}^{\mb{g} \mb{g}'}(\mb{q},\omega)}{z-\omega} \Bigg] d\omega \\
&= -\frac{1}{\pi} \int_0^\infty \Bigg[ \frac{-\text{Im} \tilde{\chi}^{-\mb{g} -\mb{g}'}(-\mb{q},\omega)}{z+\omega} + \frac{\text{Im} \tilde{\chi}^{\mb{g} \mb{g}'}(\mb{q},\omega)}{z-\omega} \Bigg] d\omega, \\
\tilde{\chi}^{-\mb{g} -\mb{g}'}(-\mb{q},z) 
&= -\frac{1}{\pi} \int_0^\infty \Bigg[ \frac{\text{Im} \tilde{\chi}^{-\mb{g} -\mb{g}'}(-\mb{q},-\omega)}{z+\omega} + \frac{\text{Im} \tilde{\chi}^{-\mb{g} -\mb{g}'}(-\mb{q},\omega)}{z-\omega} \Bigg] d\omega \\
&= -\frac{1}{\pi} \int_0^\infty \Bigg[ \frac{-\text{Im} \tilde{\chi}^{\mb{g} \mb{g}'}(\mb{q},\omega)}{z+\omega} + \frac{\text{Im} \tilde{\chi}^{-\mb{g} -\mb{g}'}(-\mb{q},\omega)}{z-\omega} \Bigg] d\omega. \\
\end{split}
\ee
Combine $\tilde{\chi}^{\mb{g} \mb{g}'}(\mb{q},z)$ and $\tilde{\chi}^{-\mb{g} -\mb{g}'}(-\mb{q},z)$,
\be
\label{Eq_chisym_TBG}
\begin{split}
\tilde{\chi}^{\mb{g} \mb{g}'}(\mb{q},z) + \tilde{\chi}^{-\mb{g} -\mb{g}'}(-\mb{q},z) &= -\frac{1}{\pi} \int_0^\infty \Big( \text{Im} \tilde{\chi}^{\mb{g} \mb{g}'}(\mb{q},\omega) + \text{Im} \tilde{\chi}^{-\mb{g} -\mb{g}'}(-\mb{q},\omega) \Big) \Big( \frac{1}{z-\omega} - \frac{1}{z+\omega} \Big) d\omega \\
&= -\frac{1}{\pi} \int_0^\infty \Big( \text{Im} \tilde{\chi}^{\mb{g} \mb{g}'}(\mb{q},\omega) + \text{Im} \tilde{\chi}^{-\mb{g} -\mb{g}'}(-\mb{q},\omega) \Big) \frac{2\omega(z_1^2-z_2^2-\omega^2)-i4z_1 z_2 \omega}{(z_1^2+z_2^2-\omega^2)^2+4z_2^2 \omega^2} d\omega,
\end{split}
\ee
where $z=z_1+i z_2$. Along the imaginary axis, the imaginary part of Eq.~(\ref{Eq_chisym_TBG}) vanishes. Therefore the integral along the real axis can be rotated to the imaginary axis:
\be
\begin{split}
\int_0^\infty \Big( \tilde{\chi}^{\mb{g} \mb{g}'}(\mb{q},i\omega) + \tilde{\chi}^{-\mb{g} -\mb{g}'}(-\mb{q},i\omega) \Big) d\omega &= \int_0^\infty \text{Re}\Big( \tilde{\chi}^{\mb{g} \mb{g}'}(\mb{q},i\omega) + \tilde{\chi}^{-\mb{g} -\mb{g}'}(-\mb{q},i\omega) \Big) d\omega \\
&= \int_0^\infty \text{Im}\Big( \tilde{\chi}^{\mb{g} \mb{g}'}(\mb{q},\omega) + \tilde{\chi}^{-\mb{g} -\mb{g}'}(-\mb{q},\omega) \Big) d\omega \\
\end{split}
\ee

Return to the xc energy, since only the diagonal in $\mb{g}$ elements of matrix $\boldsymbol{\chi}(\mb{q},i\omega;\lambda)$ is relevant in Eq.~(\ref{Eq_Exc_TBG2}), and the Coulomb matrix $\mb{V}(\mb{q})$ is diagonal in $\mb{g}$,
the exchange-correlation energy can also be written in the matrix product form,
\be
\label{Eq_Exc_TBG3}
\begin{split}
E_{xc}
&= \frac{n}{2} \sum\limits'_{\mb{q}} \Big[ -\frac{1}{\pi n} \int_0^1 d\lambda \int_0^\infty d\omega \text{Tr} \big(\mb{V}(\mb{q}) \boldsymbol{\chi}(\mb{q},i\omega;\lambda)\big) - \text{Tr}\big( \mb{V}(\mb{q})\big) \Big].
\end{split}
\ee

The coupling-constant-dependent density response function in Eq.~(\ref{Eq_Exc_TBG2}) and Eq.~(\ref{Eq_Exc_TBG3}) is approximated, within the RPA, with
\be
\label{chi_RPA_SI}
\begin{split}
\chi(\lambda) 
&= \tilde{\chi}_{_H}(1-\lambda V \tilde{\chi}_{_H})^{-1} \\
&=\tilde{\chi}_{_H} + \lambda \tilde{\chi}_{_H} V\tilde{\chi}_{_H}(1-\lambda V \tilde{\chi}_{_H})^{-1},
\end{split}
\ee
where $\tilde{\chi}_{_H}$ is the proper density response function of the self-consistent Hartree approximation. The expression of $\tilde{\chi}_{_H}$ of a specific flavor is shown in Eq.~(\ref{Eq:chiH}).

The exchange energy $E_x$ is the first order contribution in Eq.~(\ref{Eq_Exc_TBG2}) and Eq.~(\ref{Eq_Exc_TBG3}), {\it i.e.} arising from the first term in Eq.~(\ref{chi_RPA_SI})
\be
\label{Eq_Ex_TBG1}
\begin{split}
E_x &= \frac{n}{2} \sum\limits'_{\mb{q},\mb{g}} V(\mb{q}+\mb{g}) \Big[ - \frac{1}{\pi n} \int_0^\infty \tilde{\chi}_{_H}^{\mb{g} \mb{g}}(\mb{q},i\omega) d\omega - 1 \Big] \\
&= \frac{n}{2} \sum\limits'_{\mb{q}} \Big[ -\frac{1}{\pi n} \text{Tr} \big(\mb{V}(\mb{q}) \tilde{\boldsymbol{\chi}}_{_H}(\mb{q})\big) - \text{Tr}\big( \mb{V}(\mb{q})\big) \Big].
\end{split}
\ee
Using the Lindhard formula, the diagonal elements of $\tilde{\chi}_{_H}$ is 
\be
\begin{split}
\tilde{\chi}_{_H}^{\mb{g}\mb{g}}(\mb{q},i\omega) 
=& \frac{1}{A} \sum\limits_{n,m,\mb{k}} \Bigg( \frac{\varepsilon_{n\mb{k}} - \varepsilon_{m\mb{k}+\mb{q}}}{\omega^2 + (\varepsilon_{n\mb{k}} - \varepsilon_{m\mb{k}+\mb{q}})^2}
-i \frac{\omega}{\omega^2 + (\varepsilon_{n\mb{k}} - \varepsilon_{m\mb{k}+\mb{q}})^2} \Bigg) (f_{n\mb{k}}-f_{m\mb{k}+\mb{q}})
\Big| \sum\limits_{\alpha, \mb{g}_1}
z^{n}_{\alpha,\mb{g}_1}(\mb{k})
\bar{z}^{m}_{\alpha,\mb{g}_1+\mb{g}}(\mb{k}+\mb{q}) \Big|^2
\end{split}
\ee
and its real and imaginary parts are respectively
\be
\begin{split}
&\text{Re} \tilde{\chi}_{_H}^{\mb{g}\mb{g}}(\mb{q},i\omega)  = \frac{1}{A} \sum\limits_{n,m,\mb{k}} \frac{(\varepsilon_{n\mb{k}} - \varepsilon_{m\mb{k}+\mb{q}})(f_{n\mb{k}}-f_{m\mb{k}+\mb{q}})}{\omega^2 + (\varepsilon_{n\mb{k}} - \varepsilon_{m\mb{k}+\mb{q}})^2}
\Big| \sum\limits_{\alpha, \mb{g}_1}
z^{n}_{\alpha,\mb{g}_1}(\mb{k})
\bar{z}^{m}_{\alpha,\mb{g}_1+\mb{g}}(\mb{k}+\mb{q}) \Big|^2, \\
&\text{Im} \tilde{\chi}_{_H}^{\mb{g}\mb{g}}(\mb{q},i\omega)  = -\frac{1}{A} \sum\limits_{n,m,\mb{k}} \frac{\omega(f_{n\mb{k}}-f_{m\mb{k}+\mb{q}})}{\omega^2 + (\varepsilon_{n\mb{k}} - \varepsilon_{m\mb{k}+\mb{q}})^2}
\Big| \sum\limits_{\alpha, \mb{g}_1}
z^{n}_{\alpha,\mb{g}_1}(\mb{k})
\bar{z}^{m}_{\alpha,\mb{g}_1+\mb{g}}(\mb{k}+\mb{q}) \Big|^2.
\end{split}
\ee
$\varepsilon_{n\mb{k}}$, $z^n(\mb{k})$ are eigen-energies and eigenvectors of the self-consistent Hartree approximation.
Using the integration equality
\be
\int_0^\infty d \omega \frac{1}{\omega^2+a^2} = \frac{1}{a} \arctan \Big(\frac{\omega}{a} \Big) \Big|^\infty_0 = \frac{\pi}{2|a|},
\ee
the frequency integration of the real part of $\tilde{\chi}_{_H}^{\mb{g}\mb{g}}(\mb{q},i\omega)$ can be done analytically:
\be
\label{Eq_Rechi_int}
\begin{split}
\int_0^\infty \text{Re} \tilde{\chi}_{_H}^{\mb{g} \mb{g}}(\mb{q},i\omega) d\omega
=& \frac{\pi}{2A} \sum\limits_{n,m,\mb{k}} \frac{(\varepsilon_{n\mb{k}} - \varepsilon_{m\mb{k}+\mb{q}})(f_{n\mb{k}} - f_{m\mb{k}+\mb{q}})}{|\varepsilon_{n\mb{k}} - \varepsilon_{m\mb{k}+\mb{q}}|} \la n\mb{k}| e^{-i(\mb{q}+\mb{g}) \cdot \mb{r}} | m\mb{k}+\mb{q} \ra \la m\mb{k}+\mb{q}| e^{i(\mb{q}+\mb{g}) \cdot \mb{r}}|n\mb{k} \ra.
\end{split}
\ee
Rewrite the Fermi-Dirac occupation difference
\be
f_{n\mb{k}} - f_{m\mb{k}+\mb{q}}
= f_{n\mb{k}}(1-f_{m\mb{k}+\mb{q}})
-(1-f_{n\mb{k}})f_{m\mb{k}+\mb{q}}
\ee
and it is clear that
\be
\frac{\varepsilon_{n\mb{k}} - \varepsilon_{m\mb{k}+\mb{q}}}{|\varepsilon_{n\mb{k}} - \varepsilon_{m\mb{k}+\mb{q}}|}
=
\begin{cases}
-1, \text{ if } f_{n\mb{k}}=1 \text{ and } f_{m\mb{k}+\mb{q}}=0, \\
1, \text{ if } f_{n\mb{k}}=0 \text{ and } f_{m\mb{k}+\mb{q}}=1.
\end{cases}
\ee
Then Eq.~(\ref{Eq_Rechi_int}) becomes
\be
\label{Eq_Rechi_int2}
\begin{split}
\int_0^\infty \text{Re} \tilde{\chi}_{_H}^{\mb{g} \mb{g}}(\mb{q},i\omega) d\omega
=& -\frac{\pi}{2A} \sum\limits_{n,m,\mb{k}} [f_{n\mb{k}}(1-f_{m\mb{k}+\mb{q}}) + (1-f_{n\mb{k}})f_{m\mb{k}+\mb{q}}] \la n\mb{k}| e^{-i(\mb{q}+\mb{g}) \cdot \mb{r}} | m\mb{k}+\mb{q} \ra \la m\mb{k}+\mb{q}| e^{i(\mb{q}+\mb{g}) \cdot \mb{r}}|n\mb{k} \ra \\
=& -\frac{\pi}{2A} \sum\limits_{n,m,\mb{k}} (f_{n\mb{k}}+f_{m\mb{k}+\mb{q}}-2f_{n\mb{k}} f_{m\mb{k}+\mb{q}}) \la n\mb{k}| e^{-i(\mb{q}+\mb{g}) \cdot \mb{r}} | m\mb{k}+\mb{q} \ra \la m\mb{k}+\mb{q}| e^{i(\mb{q}+\mb{g}) \cdot \mb{r}}|n\mb{k} \ra. \\
\end{split}
\ee
The first two terms are simply total occupation number 
\be
\begin{split}
&\sum\limits_{n,m,\mb{k}} f_{n\mb{k}} \la n\mb{k}| e^{-i(\mb{q}+\mb{g}) \cdot \mb{r}} | m\mb{k}+\mb{q} \ra \la m\mb{k}+\mb{q}| e^{i(\mb{q}+\mb{g}) \cdot \mb{r}}|n\mb{k} \ra 
= \sum\limits_{n,\mb{k}} f_{n\mb{k}} = N, \\
&\sum\limits_{n,m,\mb{k}} f_{m\mb{k}+\mb{q}} \la n\mb{k}| e^{-i(\mb{q}+\mb{g}) \cdot \mb{r}} | m\mb{k}+\mb{q} \ra \la m\mb{k}+\mb{q}| e^{i(\mb{q}+\mb{g}) \cdot \mb{r}}|n\mb{k} \ra 
= \sum\limits_{m,\mb{k}} f_{m\mb{k}+\mb{q}} = N.
\end{split}
\ee
Equation~(\ref{Eq_Rechi_int2}) becomes
\be
\begin{split}
\int_0^\infty \text{Re} \tilde{\chi}_{_H}^{\mb{g} \mb{g}}(\mb{q},i\omega) d\omega
=& -\pi n +\frac{\pi}{A} \sum\limits_{n,m,\mb{k}} f_{n\mb{k}} f_{m\mb{k}+\mb{q}} 
\la n\mb{k}| e^{-i(\mb{q}+\mb{g}) \cdot \mb{r}} | m\mb{k}+\mb{q} \ra \la m\mb{k}+\mb{q}| e^{i(\mb{q}+\mb{g}) \cdot \mb{r}}|n\mb{k} \ra. \\
\end{split}
\ee
Substitute into the exchange energy Eq.~(\ref{Eq_Ex_TBG1}), the $-\pi n$ term above cancels exactly with the self-interacting term, and the exchange energy is simply
\be
\label{Eq_Ex_rhorho}
\begin{split}
E_x 
&= -\frac{1}{2A} \sum\limits'_{\mb{q},\mb{g}} V(\mb{q}+\mb{g})
\sum\limits_{n,m,\mb{k}} f_{n\mb{k}} f_{m\mb{k}+\mb{q}} \Big|\sum\limits_{\alpha,\mb{g}_1} z^{n}_{\alpha,\mb{g}_1}(\mb{k})
\bar{z}^m_{\alpha,\mb{g}_1+\mb{g}}(\mb{k}+\mb{q}) \Big|^2 \\
&= -\frac{1}{2A} \sum\limits'_{\mb{q},\mb{g}} V(\mb{q}+\mb{g})
\sum\limits_{n,m,\mb{k}} f_{n\mb{k}} f_{m\mb{k}+\mb{q}} \sum\limits_{\alpha,\beta,\mb{g}_1,\mb{g}_2} \bar{\rho}^{n}_{\alpha,\mb{g}_1;\beta,\mb{g}_2}(\mb{k})
\rho^m_{\alpha,\mb{g}_1+\mb{g};\beta,\mb{g}_2+\mb{g}}(\mb{k}+\mb{q}).
\end{split}
\ee
This exchange energy looks like the HF exchange but it is not since it is calculated over the SCH wavefunctions instead of the self-consistent HF wavefunctions.

On the other hand, the correlation energy is given by higher order contributions in RPA. Use
\be
\text{Tr} (\mb{V}\boldsymbol{\chi}) = \text{Tr} (\sqrt{\mb{V}} \boldsymbol{\chi} \sqrt{\mb{V}}),
\ee
the correlation energy is
\be
\begin{split}
E_c
&= -\frac{1}{2\pi} \sum\limits'_{\mb{q}} \int_0^1 d\lambda \int_0^\infty d\omega
\text{Tr} \big[\lambda (\sqrt{\mb{V}} \tilde{\bs{\chi}}_{_H} \sqrt{\mb{V}})^2(1-\lambda \sqrt{\mb{V}} \tilde{\bs{\chi}}_{_H} \sqrt{\mb{V}})^{-1} \big]\\
&= \frac{1}{2\pi} \sum\limits'_{\mb{q}}
\int_0^\infty d\omega \text{Tr} \big[ \sqrt{\mb{V}} \tilde{\bs{\chi}}_{_H} \sqrt{\mb{V}} + \ln (1- \sqrt{\mb{V}} \tilde{\bs{\chi}}_{_H} \sqrt{\mb{V}}) \big].
\end{split}
\ee
{For any diagonalizable and non-singular (invertible) matrix $\mb{A}$, its logarithm is
\be
\label{Eq_lnA}
\ln \mb{A} = \mb{v} (\ln \bs{\lambda}) \mb{v}^{-1},
\ee
where $\bs{\lambda}$ is the diagonal matrix of eigenvalues of $\mb{A}$,
\be
\bs{\lambda} = \mb{v}^{-1} \mb{A} \mb{v},
\ee
and $\mb{v}$ is the matrix with eigenvectors in each column. 
The trace of $\ln \mb{A}$
\be
\begin{split}
\text{Tr} \ln \mb{A} 
&= \text{Tr} \big[ \mb{v} (\ln \boldsymbol{\lambda}) \mb{v}^{-1} \big] \\
&= \text{Tr} \big[(\ln \boldsymbol{\lambda}) \mb{v}^{-1} \mb{v} \big] \\
&= \text{Tr} \ln \boldsymbol{\lambda} \\
&= \ln \det \mb{A}.
\end{split}
\ee
The correlation energy 
\be
\begin{split}
E_c
&= \frac{1}{2\pi} \sum\limits'_{\mb{q}} \int_0^\infty d\omega \text{Tr} \Big[
\mb{V}(\mb{q}) \tilde{\bs{\chi}}_{_H}(\mb{q},i\omega) + \ln \big(1- \sqrt{\mb{V}(\mb{q})} \tilde{\bs{\chi}}_{_H}(\mb{q},i\omega) \sqrt{\mb{V}(\mb{q})} \big) \Big]\\
&= \frac{1}{2\pi} \sum\limits'_{\mb{q}} \int_0^\infty d\omega \Big[
\text{Tr} \big( \mb{V}(\mb{q}) \tilde{\bs{\chi}}_{_H}(\mb{q},i\omega) \big) + \sum\limits_i \ln \lambda_i(\mb{q},i\omega) \Big],
\end{split}
\ee
where $\lambda_i(\mb{q},i\omega)$ is the $i$-th eigenvalue of $(1- \sqrt{\mb{V}} \tilde{\bs{\chi}}_{_H} \sqrt{\mb{V}})$.

\section{Exchange energy regularization}\label{SM:Ex_regularization}
The exchange energy in Eq.~(\ref{Eq_Ex_rhorho}) must be regularized to deal with the negative energy sea of the Dirac model. In the main text we denote $E_x$ as the regularized exchange energy
\be
\label{Eq_Ex_reg}
\begin{split}
E_x 
&= -\frac{1}{2A} \sum\limits'_{\mb{q},\mb{g}} V(\mb{q}+\mb{g}) \sum\limits_{\substack{\mb{k},\mb{g}_1,\mb{g}_2 \\ \alpha, \beta}}
\big[ \bar{\rho}_{\alpha,\mb{g}_1;\beta,\mb{g}_2}(\mb{k})
\rho_{\alpha,\mb{g}_1+\mb{g};\beta,\mb{g}_2+\mb{g}}(\mb{k}+\mb{q}) - \bar{\rho}^0_{\alpha,\mb{g}_1;\beta,\mb{g}_2}(\mb{k})
\rho^0_{\alpha,\mb{g}_1+\mb{g};\beta,\mb{g}_2+\mb{g}}(\mb{k}+\mb{q}) \big] \\
&= -\frac{1}{2A} \sum\limits'_{\mb{q},\mb{g}} V(\mb{q}+\mb{g}) \sum\limits_{\substack{\mb{k},\mb{g}_1,\mb{g}_2 \\ \alpha, \beta}} \big[\delta \bar{\rho}(\mb{k}) + 2 \bar{\rho}^{0}(\mb{k}) \big]_{\alpha,\mb{g}_1;\beta,\mb{g}_2} \delta \rho_{\alpha, \mb{g}_1+\mb{g};\beta,\mb{g}_2+\mb{g}}(\mb{k}+\mb{q}).
\end{split}
\ee
$\rho^0$ is the density matrix of the charge neutral decoupled bilayer and therefore it's diagonal in $\mb{g}$'s: 
\be
\rho^0_{\alpha,\mb{g}_1;\beta,\mb{g}_2} (\mb{k}) = \delta_{\mb{g}_1 \mb{g}_2} \sum\limits_{n \in v} z^0_{n,\alpha,\mb{g}_1}(\mb{k}) \bar{z}^0_{n,\beta,\mb{g}_2}(\mb{k}).
\ee
Summing over two valence bands from top and bottom layer Dirac cones, the $4 \times 4$ density matrix with an explicit $\mb{g}$ is
\be
\rho^0_{\mb{g};\mb{g}}(\mb{k}) \equiv \rho^0(\mb{k}+\mb{g}) = \frac{1}{2}
\begin{pmatrix}
1 & -e^{-i(\theta_{\mathbf{k}+\mathbf{g}-\mathbf{K}_1}-\theta/2)} & 0 & 0 \\
-e^{i(\theta_{\mathbf{k}+\mathbf{g}-\mathbf{K}_1}-\theta/2)} & 1 & 0 & 0 \\
0 & 0 & 1 & -e^{-i(\theta_{\mathbf{k}+\mathbf{g}-\mathbf{K}_2}+\theta/2)} \\
0 & 0 & -e^{i(\theta_{\mathbf{k}+\mathbf{g}-\mathbf{K}_2}+\theta/2)} & 1
\end{pmatrix}.
\ee
Therefore, the second part of the second line of Eq.~(\ref{Eq_Ex_reg}) is equivalent to
\be
\begin{split}
E_x^{(2)} 
&= -\frac{1}{A} \sum\limits_{\substack{\mathbf{k}' \in \text{MBZ} \\
\mathbf{k} \in k_c \sim 1/a}} V(\mathbf{k}'-\mathbf{k}+\mb{g}) 
\sum\limits_{\mb{g}_1,\alpha,\beta}
\bar{\rho}^0_{\alpha \beta}(\mathbf{k}+\mb{g}_1) \delta \rho_{\alpha,\mb{g}_1+\mb{g};\beta,\mb{g}_1+\mb{g}}(\mathbf{k}') 
\end{split}
\ee
Because the diagonal terms of $\rho^0$ contribute a constant energy shift which is proportional to the occupation number at $\mb{k}'$, only off-diagonal terms of $\rho^0$ matter and it's
\be
\begin{split}
E_x^{(2)}
= \frac{2\pi e^2}{\epsilon}\frac{1}{8\pi} \sum\limits_{\mathbf{k}' \in \text{MBZ},\mathbf{g}} \Big[ & e^{i\xi(\theta_{\mathbf{k}'+\mathbf{g}-\mathbf{K}_1}-\theta/2)} [\delta \rho_{\mathbf{g};\mathbf{g}}]_{12}(\mathbf{k}') |\mathbf{k}'+\mathbf{g}-\mathbf{K}_1| \ln \frac{k_c}{|\mathbf{k}'+\mathbf{g}-\mathbf{K}_1|} \\
+& e^{i\xi(\theta_{\mathbf{k}'+\mathbf{g}-\mathbf{K}_2}+\theta/2)} \delta [\rho_{\mathbf{g};\mathbf{g}}]_{34}(\mathbf{k}') |\mathbf{k}'+\mathbf{g}-\mathbf{K}_2| \ln \frac{k_c}{|\mathbf{k}'+\mathbf{g}-\mathbf{K}_2|} + h.c \Big]
\end{split}
\ee

Therefore the regularized exchange energy is
\be
\label{Eq_Ex_reg2}
\begin{split}
E_x 
&= -\frac{1}{2A} \sum\limits'_{\mb{q},\mb{g}} V(\mb{q}+\mb{g}) \sum\limits_{\substack{\mb{k},\mb{g}_1,\mb{g}_2 \\ \alpha, \beta}} \delta \bar{\rho}_{\alpha,\mb{g}_1;\beta,\mb{g}_2} (\mb{k}) \delta \rho_{\alpha, \mb{g}_1+\mb{g};\beta,\mb{g}_2+\mb{g}}(\mb{k}+\mb{q}) + E^{(2)}_x.
\end{split}
\ee

\section{Screening effects from remote bands}
\label{SM:remotebandscreening}
In Fig.~\ref{Fig_Vchi_vs_nu}(a) we show the density of state (DOS) and $V_{\mb{q}+\mb{g}} \tilde{\chi}_{_H}^{\mb{g} \mb{g}}(\mb{q},i\omega=0)$ as a function of $\nu$, for the specific $\mb{g}$ and $\mb{q}$. For both flavor paramagnetic state (solid lines) and flavor polarized state (dashed lines), electron-hole excitations between flat bands (ff) dominate over excitations between remote and flat bands (rf) and between remote bands (rr), as long as flat bands are not entirely empty or fully occupied. In Fig.~\ref{Fig_Vchi_vs_nu}(b), contributions to the correlation energy ($E_c$) from electron-hole excitations between flat bands ($E_c^{\text{ff}}$), between remote and flat bands ($E_c^{\text{rf}}$) and between remote bands ($E_c^{\text{rr}}$) are separately shown. $E_c^{\text{ff}}$ dominates and is responsible for the tendency of $E_c$ with respect to $\nu$. By comparing correlation energies of flavor paramagnetic state (solid lines) with flavor polarized state (dashed lines), $E_c^{\text{rr}}$ and $E_c^{\text{rf}}$ are almost independent of flavor polarizations.

Figure~\ref{Fig:chiHwrtOmega} shows $V_{\mb{q}+\mb{g}} \tilde{\chi}_{_H}^{\mb{g} \mb{g}}(\mb{q},i\omega)$ at $\nu=-1$ as a function of frequency, for the specific $\mb{g}$ and $\mb{q}$. It illustrates that the xc energy are dominated by diagonal elements of $\mb{V} \tilde{\boldsymbol{\chi}}_{_H}$ with the smallest $|\mb{g}|$ and by excitations between flat bands (ff).

Figure~\ref{Fig:valchi_wrtqOmega} shows the sum of eigenvalues of $\ln(1-\mb{V} \tilde{\boldsymbol{\chi}})$, {\it{i.e.}} the second term of Eq.~(\ref{Eq:Ec}), as a function of $q$ and unitless frequency $\tilde{\omega}$. Again the correlation effect is dominated by ff excitations.

\begin{figure}
  \includegraphics[width=0.8\columnwidth]{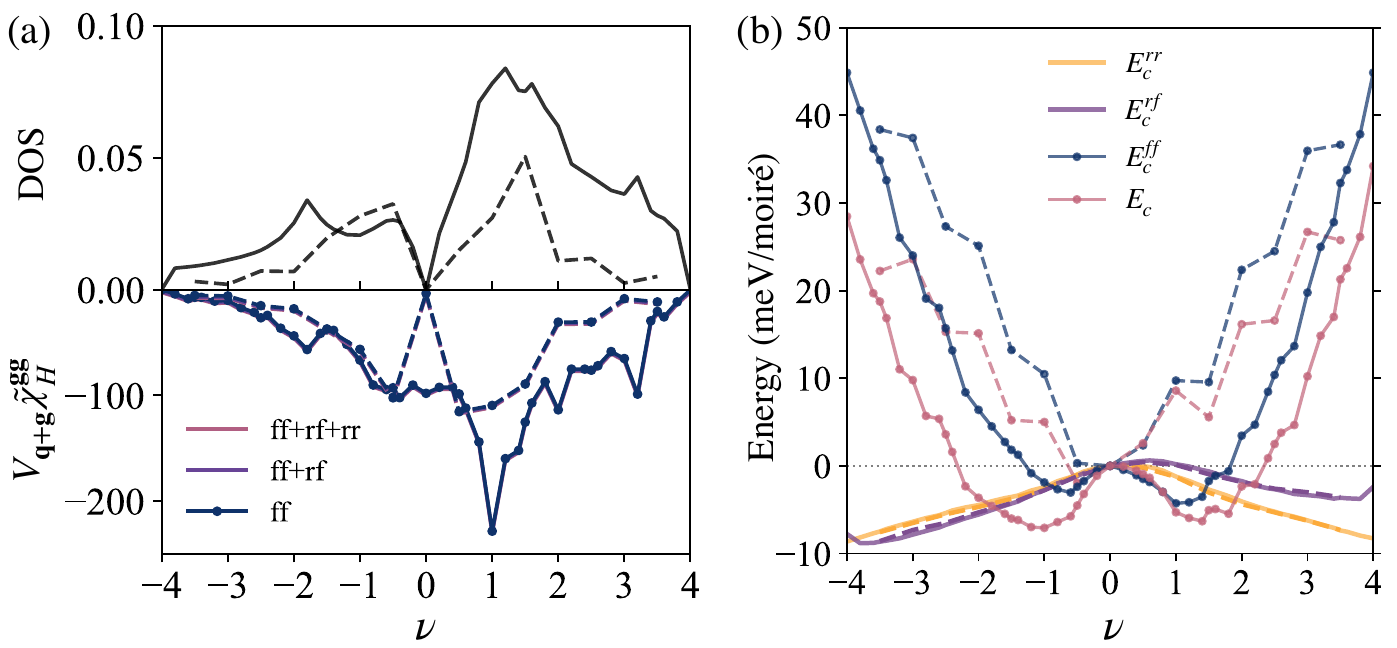}
  \vspace{-10pt}
  \caption{\small (a) Density of state in unit of meV$^{-1} \cdot$nm$^{-2}$ and $V_{\mb{q}+\mb{g}} \tilde{\chi}_{_H}^{\mb{g} \mb{g}}(\mb{q
  }, i\omega=0)$ as a function of $\nu$, with $\mb{g}=0$ and $\mb{q} = (-\sqrt{3}/2, -1/2) g_{\rm M}/\sqrt{3}$, where $g_{\rm M}$ is the length of moir\'e primitive reciprocal lattice vector. Different colors in $V_{\mb{q}+\mb{g}} \tilde{\chi}_{_H}^{\mb{g} \mb{g}}(\mb{q
  }, i\omega=0)$ plot label distinct particle-hole excitations as indicated by the legend. (a) shows that excitations between flat bands (ff) dominate when the flat bands are not empty or fully occupied.
  (b) The correlation energy $E_c$ as a function of $\nu$. $E_c^{\text{ff}}$, $E_c^{\text{rf}}$ and $E_c^{\text{rr}}$ are contributions to the correlation energy from excitations between flat bands, between remote bands and flat bands and between remote bands respectively. $E_c^{\text{rr}}$ and $E_c^{\text{rf}}$ are independent of flavor polarizations. In both figures, solid lines represent flavor paramagnetic states and dashed lines represent flavor fully polarized states, for example at $\nu=2$, $\nu_f=(1,1,0,0)$. 
  }
  \label{Fig_Vchi_vs_nu}
\end{figure}

\begin{figure}
  \includegraphics[width=0.8\columnwidth]{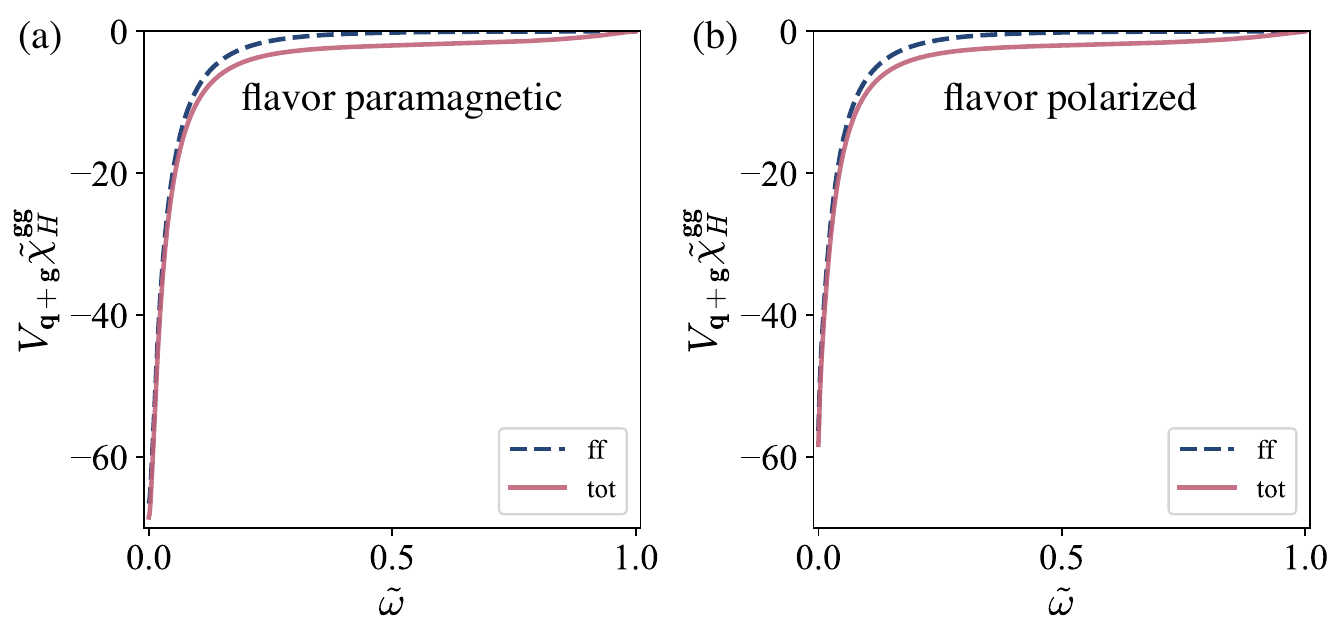}
  \vspace{-10pt}
  \caption{\small $V(\mb{q}+\mb{g}) \tilde{\chi}_{_H}^{\mb{g} \mb{g}}(\mb{q
  }, i\omega)$ at $\nu=-1$ as a function of the unitless frequency $\tilde{\omega}$, with $\mb{g}=0$ and $\mb{q} = (-\sqrt{3}/2, -1/2) g_{\rm M}/\sqrt{3}$. 
  $\tilde{\omega}=\omega/(\omega+\omega_0)$, where $\omega_0=30$ meV is chosen to be around the size of flat-band bandwidth.
  Electron-hole excitations between flat bands (ff) play a dominant role in the correlation effect.
  }
  \label{Fig:chiHwrtOmega}
\end{figure}

\begin{figure}
  \includegraphics[width=0.8\columnwidth]{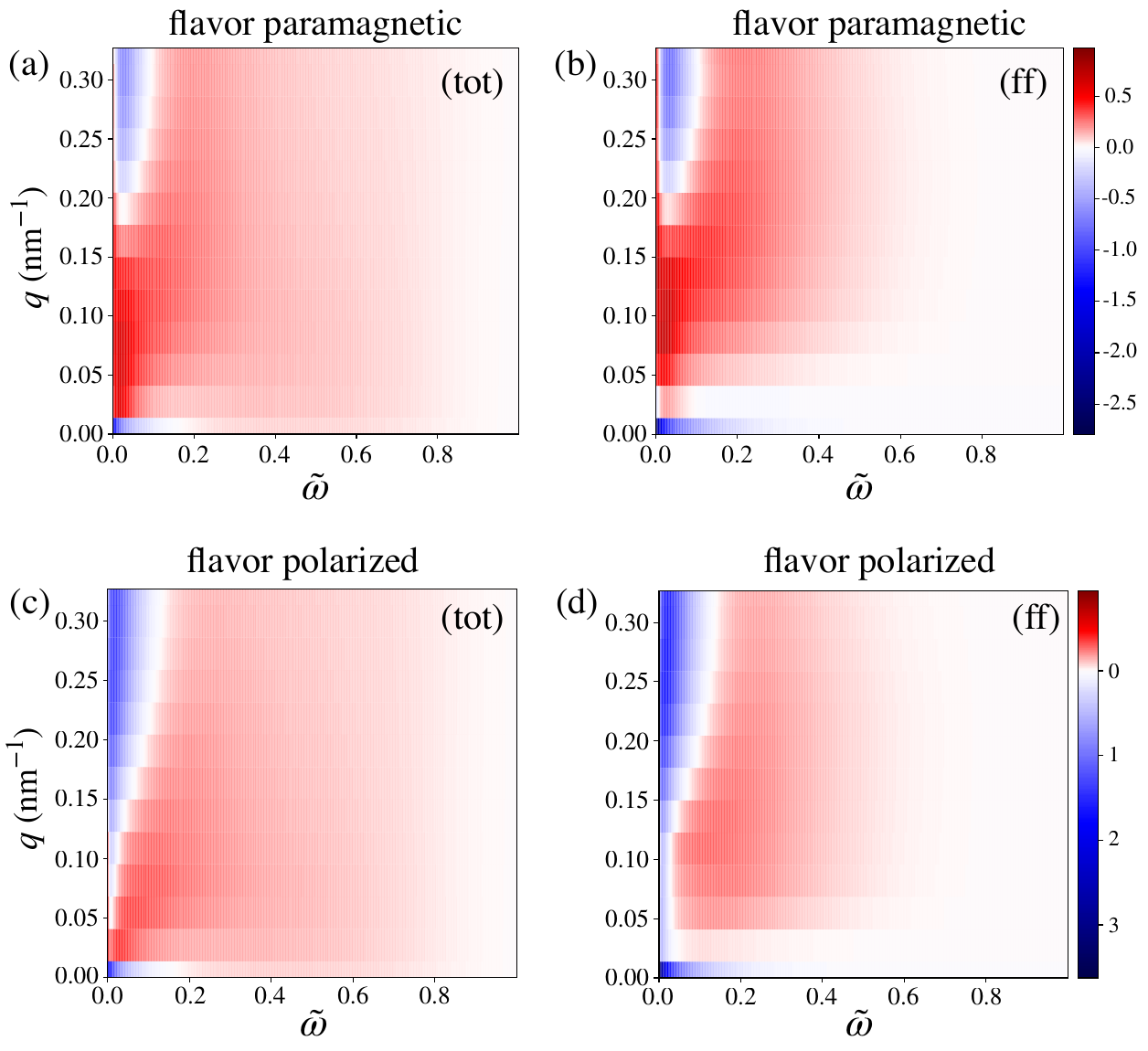}
  \vspace{-10pt}
  \caption{\small $\sum\limits_j \ln(1-\lambda_j)$ with respect to $q$ and unitless frequency $\tilde{\omega}$, where $\lambda_j$ is the $j$-th eigenvalue of $\mb{V} \tilde{\boldsymbol{\chi}}_{_H}$.
  }
  \label{Fig:valchi_wrtqOmega}
\end{figure}

\section{Energies of competing broken flavor-symmetry states}
\label{SM:Table_energy}

\begin{table*}[t]
\resizebox{\dimexpr \columnwidth}{!}{
\begin{tabular}{ |c|c|c|c|c|c|c||c|c|c|c|c|c|c| }
\hline
$\nu$ & ($\nu_1$, $\nu_2$, $\nu_3$, $\nu_4$) & $E_0$ & $E_x$ & $E_c$ & $E_{xc}$ & $E_{tot}$ & $\nu$ & ($\nu_1$, $\nu_2$, $\nu_3$, $\nu_4$) & $E_0$ & $E_x$ & $E_c$ & $E_{xc}$ & $E_{tot}$ \\
\hline
\multirow{5}{2em}{\centering -0.5} & (-1/8, -1/8, -1/8, -1/8) & -5.477 & 6.097 & -4.520 & 1.577 & -3.900 & \multirow{5}{2em}{\centering 0.5} & (1/8, 1/8, 1/8, 1/8) & 6.852 & 1.570 & -0.930 & \bf{0.640} & \bf{7.492} \\
& (-1/4, -1/8, -1/8, 0) & -5.458 & 5.534 & -3.992 & \bf{1.542} & \bf{-3.916} & & 
(1/4, 1/8, 1/8, 0) & 6.864 & 1.163 & -0.443 & 0.720 & 7.584 \\
& (-1/4, -1/4, 0, 0) & -5.437 & 5.417 & -3.834 & 1.583 & -3.854 & & (1/4, 1/4, 0, 0) & 6.876 & 0.756 & 0.033 & 0.789 & 7.665 \\
& (-3/8, -1/8, 0, 0) & -5.426 & 4.650 & -3.039 & 1.611 & -3.815 & & 
(3/8, 1/8, 0, 0) & 6.882 & 0.006 & 0.702 & 0.708 & 7.590 \\
& (-1/2, 0, 0, 0) & -5.386 & 3.586 & -1.913 & 1.673 & -3.713 & & (1/2, 0, 0, 0) & 6.905 & -1.863 & 2.579 & 0.716 & 7.621 \\
\hline
\multirow{5}{2em}{\centering -1.0} & (-1/4, -1/4, -1/4, -1/4) & -10.236 & 11.187 & -7.073 & 4.114 & \bf{-6.122} & \multirow{5}{2em}{\centering 1.0} & (1/4, 1/4, 1/4, 1/4) & 14.774 & 8.384 & -5.293 & 3.091 & 17.865 \\
& (-1/2, -1/4, -1/4, 0) & -10.138 & 9.902 & -5.659 & 4.243 & -5.895 & & (1/2, 1/4, 1/4, 0) & 14.806 & 6.313 & -2.983 & 3.330 & 18.136 \\
& (-1/2, -1/2, 0, 0) & -10.044 & 8.465 & -4.162 & 4.303 & -5.741 & & (1/2, 1/2, 0, 0) & 14.833 & 4.721 & -1.500 & 3.221 & 18.054 \\
& (-3/4, -1/4, 0, 0) & -10.006 & 6.076 & -1.946 & 4.130 & -5.876 & & (3/4, 1/4, 0, 0) & 14.848 & 1.755 & 1.349 & 3.104 & 17.952 \\
& (-1, 0, 0, 0) & -9.816 & -0.957 & 5.008 & \bf{4.051} & -5.765 & & (1, 0, 0, 0) & 14.907 & -5.881 & 8.586 & \bf{2.705} & \bf{17.612} \\
\hline
\multirow{6}{2em}{\centering -1.5} & (-3/8, -3/8, -3/8, -3/8) & -13.764 & 11.482 & -6.009 & 5.473 & \bf{-8.291} & \multirow{6}{2em}{\centering 1.5} & 
(3/8, 3/8, 3/8, 3/8) & 23.162 & 11.808 & -5.054 & 6.754 & 29.916 \\
& (-3/4, -3/8, -3/8, 0) & -13.559 & 7.287 & -1.825 & 5.462 & -8.097 & & 
(3/4, 3/8, 3/8, 0) & 23.239 & 7.646 & -1.159 & 6.487 & 29.726 \\
& (-3/4, -3/4, 0, 0) & -13.335 & 3.008 & 2.418 & 5.426 & -7.909 & & 
(3/4, 3/4, 0, 0) & 23.328 & 3.450 & 2.815 & 6.265 & 29.593 \\
& (-1, -1/6, -1/6, -1/6) & -13.450 & 1.424 & 3.749 & \bf{5.173} & -8.277 & & 
(1, 1/6, 1/6, 1/6) & 23.261 & 2.345 & 3.882 & 6.227 & 29.488 \\
& (-1, -1/4, -1/4, 0) & -13.410 & 1.267 & 4.071 & 5.338 & -8.072 & & 
(1, 1/4, 1/4, 0) & 23.297 & 1.984 & 4.065 & \bf{6.049} & \bf{29.346} \\
& (-1, -1/2, 0, 0) & -13.266 & 0.215 & 5.212 & 5.427 & -7.839 & & 
(1, 1/2, 0, 0) & 23.355 & 0.570 & 5.556 & 6.126 & 29.481 \\
\hline
\multirow{4}{2em}{\centering -2.0} & (-1/2, -1/2, -1/2, -1/2) & -15.756 & 9.684 & -3.635 & 6.049 & \bf{-9.707} & \multirow{4}{2em}{\centering 2.0} & 
(1/2, 1/2, 1/2, 1/2) & 32.588 & 11.054 & -2.341 & 8.713 & 41.301 \\
& (-1, -1/3, -1/3, -1/3) & -15.530 & 1.340 & 4.537 & 5.877 & -9.653 & & 
(1, 1/3, 1/3, 1/3) & 32.681 & 4.395 & 3.984 & 8.379 & 41.060 \\
& (-1, -1/2, -1/2, 0) & -15.314 & 0.227 & 5.721 & 5.948 & -9.366 & & (1, 1/2, 1/2, 0) & 32.793 & 1.151 & 6.991 & 8.142 & 40.935 \\
& (-1, -1, 0, 0) & -14.862 & -9.434 & 15.127 & \bf{5.693} & -9.169 & & (1, 1, 0, 0) & 32.985 & -8.577 & 16.158 & \bf{7.581} & \bf{40.566} \\
\hline
\multirow{7}{2em}{\centering -2.5} & (-5/8, -5/8, -5/8, -5/8) & -16.095 & 2.991 & 3.595 & 6.586 & -9.509 & \multirow{7}{2em}{\centering 2.5} & (5/8, 5/8, 5/8, 5/8) & 43.627 & 7.017 & 2.494 & 9.511 & 53.138 \\
& (-1, -1/2, -1/2, -1/2) & -15.919 & 0.494 & 5.962 & 6.456 & -9.463 & & 
(1, 1/2, 1/2, 1/2) & 43.685 & 3.354 & 6.068 & 9.422 & 53.107 \\
& (-1, -3/4, -3/8, -3/8) & -15.844 & -2.101 & 8.436 & 6.335 & \bf{-9.509} & & 
(1, 3/4, 3/8, 3/8) & 43.734 & 0.136 & 9.266 & 9.402 & 53.136 \\
& (-1, -3/4, -2/4, -1/4) & -15.812 & -2.242 & 8.691 & 6.449 & -9.363 & & 
(1, 3/4, 2/4, 1/4) & 43.746 & -0.106 & 9.424 & 9.318 & 53.064 \\
& (-1, -3/4, -3/4, 0,) & -15.483 & -6.518 & 12.820 & 6.302 & -9.181 & & (1, 3/4, 3/4, 0) & 43.925 & -4.951 & 13.961 & 9.010 & 52.935 \\
& (-1, -1, -1/4, -1/4) & -15.582 & -7.584 & 13.887 & 6.303 & -9.279 & & (1, 1, 1/4, 1/4) & 43.845 & -6.466 & 15.462 & 8.996 & 52.841 \\
& (-1, -1, -1/2, 0) & -15.361 & -9.125 & 15.295 & \bf{6.170} & -9.191 & & (1, 1, 1/2, 0) & 43.973 & -7.790 & 16.593 & \bf{8.803} & \bf{52.776} \\
\hline
\multirow{5}{2em}{\centering -3.0} & (-3/4, -3/4, -3/4, -3/4) & -14.731 & -2.788 & 9.788 & 7.000 & -7.731 & \multirow{5}{2em}{\centering 3.0} & (3/4, 3/4, 3/4, 3/4) & 56.420 &  0.058 & 10.219 & 10.277 & 66.697 \\
& (-1, -2/3, -2/3, -2/3) & -14.627 & -5.009 & 11.905 & 6.896 & -7.731 & & 
(1, 2/3, 2/3, 2/3) & 56.443 & -1.905 & 12.109 & 10.204 & 66.647 \\
& (-1, -1, -1/2, -1/2) & -14.425 & -7.827 & 14.517 & 6.690 & \bf{-7.735} & & (1, 1, 1/2, 1/2) & 56.523 & -6.657 & 16.577 & 9.920 & 66.443 \\
& (-1, -1, -3/4, -1/4) & -14.301 & -10.579 & 17.277 & 6.698 & -7.603 & & (1, 1, 3/4, 1/4) & 56.585 & -9.340 & 19.149 & 9.809 & 66.394 \\
& (-1, -1, -1, 0) & -13.748 & -17.295 & 23.588 & \bf{6.293} & -7.455 & & (1, 1, 1, 0) & 56.897 & -17.628 & 26.698 & \bf{9.070} & \bf{65.967} \\
\hline
\multirow{4}{2em}{\centering -3.5} & (-7/8, -7/8, -7/8, -7/8) & -11.595 & -11.455 & 18.756 & 7.301 & -4.294 & \multirow{4}{2em}{\centering 3.5} & (7/8, 7/8, 7/8, 7/8) & 71.002 & -10.902 & 21.286 & 10.384 & 81.386 \\
& (-1, -5/6, -5/6, -5/6) & -11.566 & -11.142 & 18.420 & 7.278 & -4.288 & & 
(1, 5/6, 5/6, 5/6) & 71.014 & -10.676 & 21.261 & 10.585 & 81.599 \\
& (-1, -1, -3/4, -3/4) & -11.501 & -12.790 & 19.905 & 7.115 & \bf{-4.386} & & 
(1, 1, 3/4, 3/4) & 71.028 & -12.365 & 22.870 & 10.505 & 81.533 \\
& (-1, -1, -1, -1/2) & -11.322 & -15.305 & 22.244 & \bf{6.939} & -4.383 & & 
(1, 1, 1, 1/2) & 71.103 & -15.514 & 25.758 & \bf{10.244} & \bf{81.347} \\
\hline
\end{tabular}}
\caption{
\label{tab:energy}
The exchange $E_x$, the correlation $E_c$ and the total $E_{tot}$ energies of competing flavor-symmetry broken states, calculated using $\epsilon_{\text{BN}}=5.1$. The lowest $E_{tot}$ and $E_{xc}$ are marked bold at each $\nu$. On the electron-doped side, the ground state prefers flavor paramagnetism for $|\nu| < 1.0$ and flavor polarization for $|\nu| \gtrsim 1.0$. On the hole-doped side, however, the ground state favors flavor paramagnetism for $|\nu| \lesssim 2.0$ and flavor polarization for $|\nu| \gtrsim 2.0$.
Energies are in the unit of meV per moir\'e unit cell. The kinetic energy $E_k$, defined as $E_k = E_{band} - 2E_H$ where $E_{band}$ and $E_H$ are the band energy and the Hartree energy of the SCH quasi-particle bands respectively, is regularized by the kinetic energy of flavor paramagnetic state at each filling $\nu$.}
\end{table*}

Figure~\ref{Fig:Exc_polarize} schematically summarizes our findings in Table~\ref{tab:energy} by plotting the total energy (Fig.~\ref{Fig:Exc_polarize}(a)) and xc energy (Fig.~\ref{Fig:Exc_polarize}(b)) relative to the flavor paramagnetic state for various flavor polarizations ($y$-axis) as a function of filling factor $\nu$ ($x$-axis). $E^S_{tot}-E_{tot}>0$ indicates the ground state favors flavor polarized states. $E^S_{xc}-E_{xc}>0$ indicates the xc effect, when the single-quasiparticle energy is ignored, favors flavor polarized states. Figure ~\ref{Fig:Exc_polarize}(b) clearly shows that the xc effect predicts broken flavor symmetry for $|\nu| \geq 1$ on both electron- and hole-doped sides and is stronger on the electron-doped side. After including the single-quasiparticle energy which in general prefers unpolarized state, however, the ground state is predicted to be flavor paramagnetic on hole-doped side. This is a result of the fact that on the hole-doped side the Hartree energy is twice stronger and the xc effect is weaker than that on the electron-doped side.

We further explore the effects of $C_2T$ symmetry breaking on energies, as summarized in Table~\ref{tab:energy_woC2T}. We find that flavor ferromagnetic states are favored at all filling factors including those near CN, consistent with experiments that hBN alignment favors broken symmetry states.

\begin{figure}
  \includegraphics[width=0.6\columnwidth]{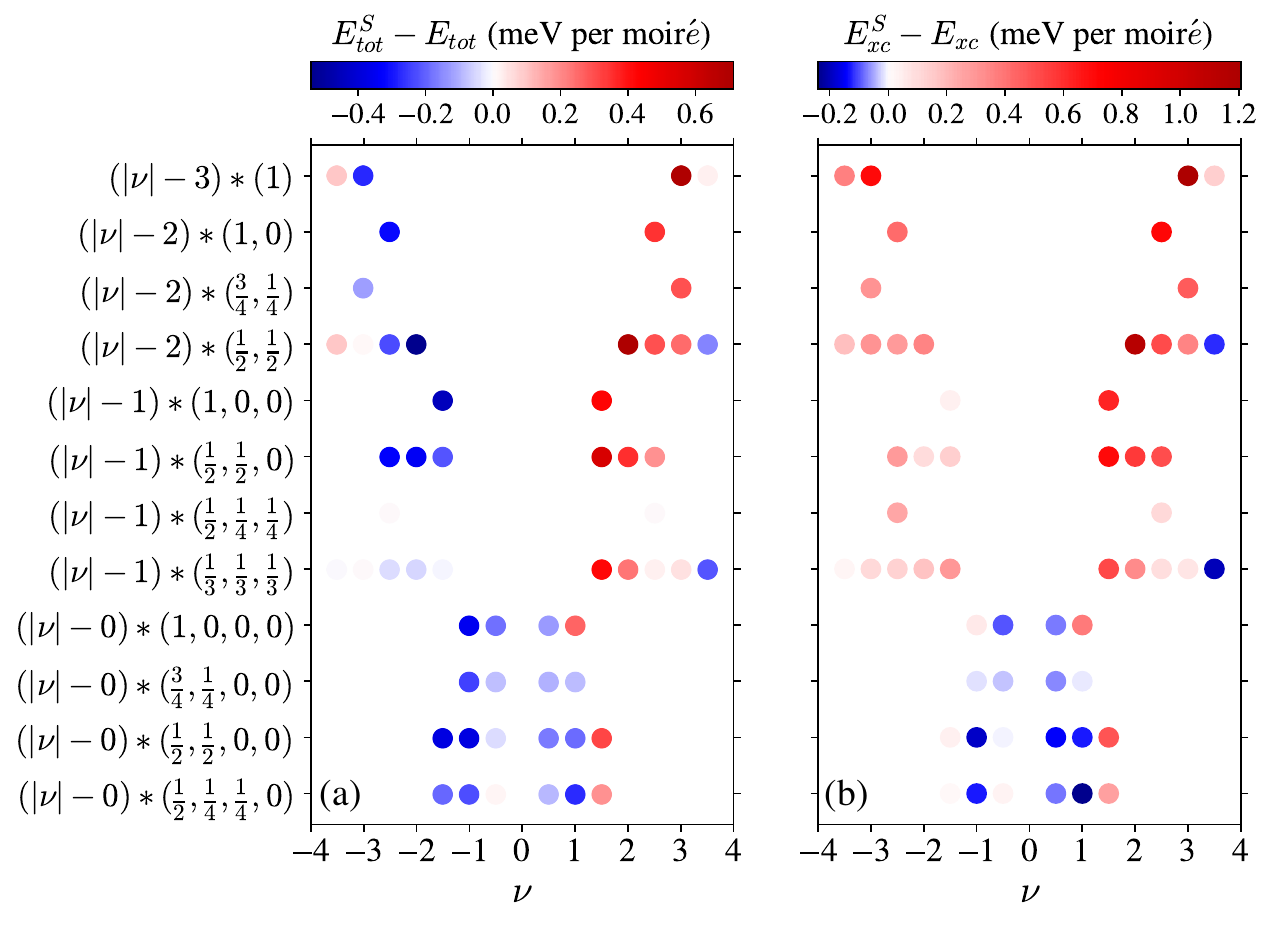}
  \vspace{-10pt}
  \caption{\small (a) Total energy $E_{tot}$ and (b) xc energy ($E_{xc}$) relative to those of the flavor paramagnetic state ($E_{tot}^S$, $E_{xc}^S$), for various flavor-polarized states ($y$-axis) as a function of $\nu$ ($x$-axis). The $y$-axis is labeled by $(|\nu|-n)*(f_1, ...)$, where the integer $n$ indicates the number of flavors that are fully occupied or empty and fractions inside the right bracket, multiplied by $(|\nu|-n)$, denote filling factors of the remaining partially occupied flavors. 
  For example, the data point at $\nu=-1$ and $(|\nu|-0)*(1/2, 1/4, 1/4, 0)$ represents the filling factors of four flavors are $-(1/2, 1/4, 1/4, 0)$ respectively.
  $E^S_{tot}-E_{tot}>0$ indicates the ground state favors flavor polarized states and $E^S_{xc}-E_{xc}>0$ indicates the xc energy favors flavor polarized states.
  }
  \label{Fig:Exc_polarize}
\end{figure}

\begin{figure}
  \includegraphics[width=0.9\columnwidth]{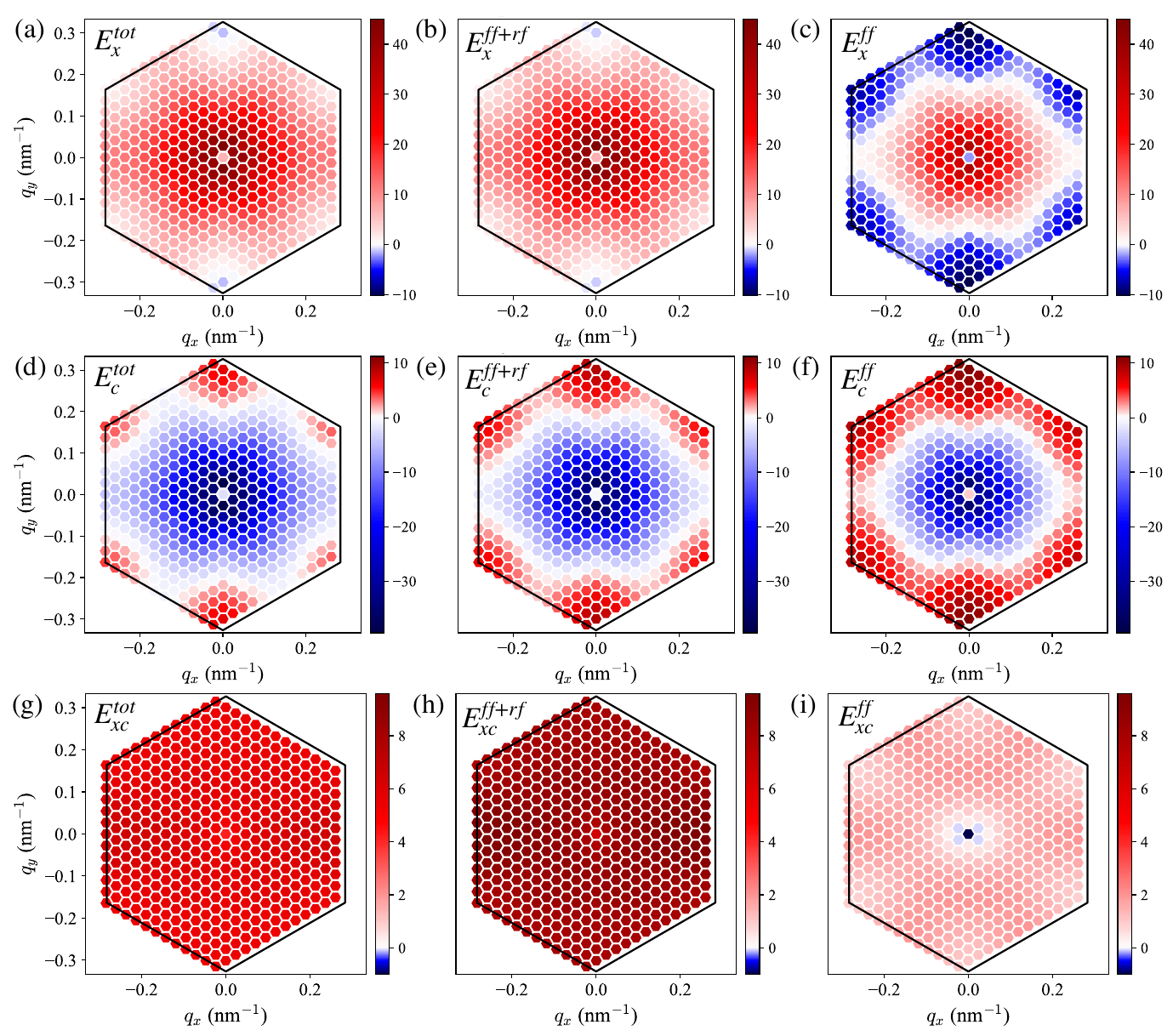}
  \vspace{-10pt}
  \caption{\small Energies versus $\mb{q}$ of the paramagnetic state  at $\nu=-1$. (a-c) The exchange energy including excitations between all SCH bands ($E_x^{\text{tot}}$), between flat bands and between flat and remote bands ($E_x^{\text{ff+rf}}$), between flat bands only ($E_x^{\text{ff}}$). (d-f) The correlation energy. (g-i) The xc energy.
  }
  \label{Fig:E_vs_q_1111}
\end{figure}

\begin{figure}
  \includegraphics[width=0.9\columnwidth]{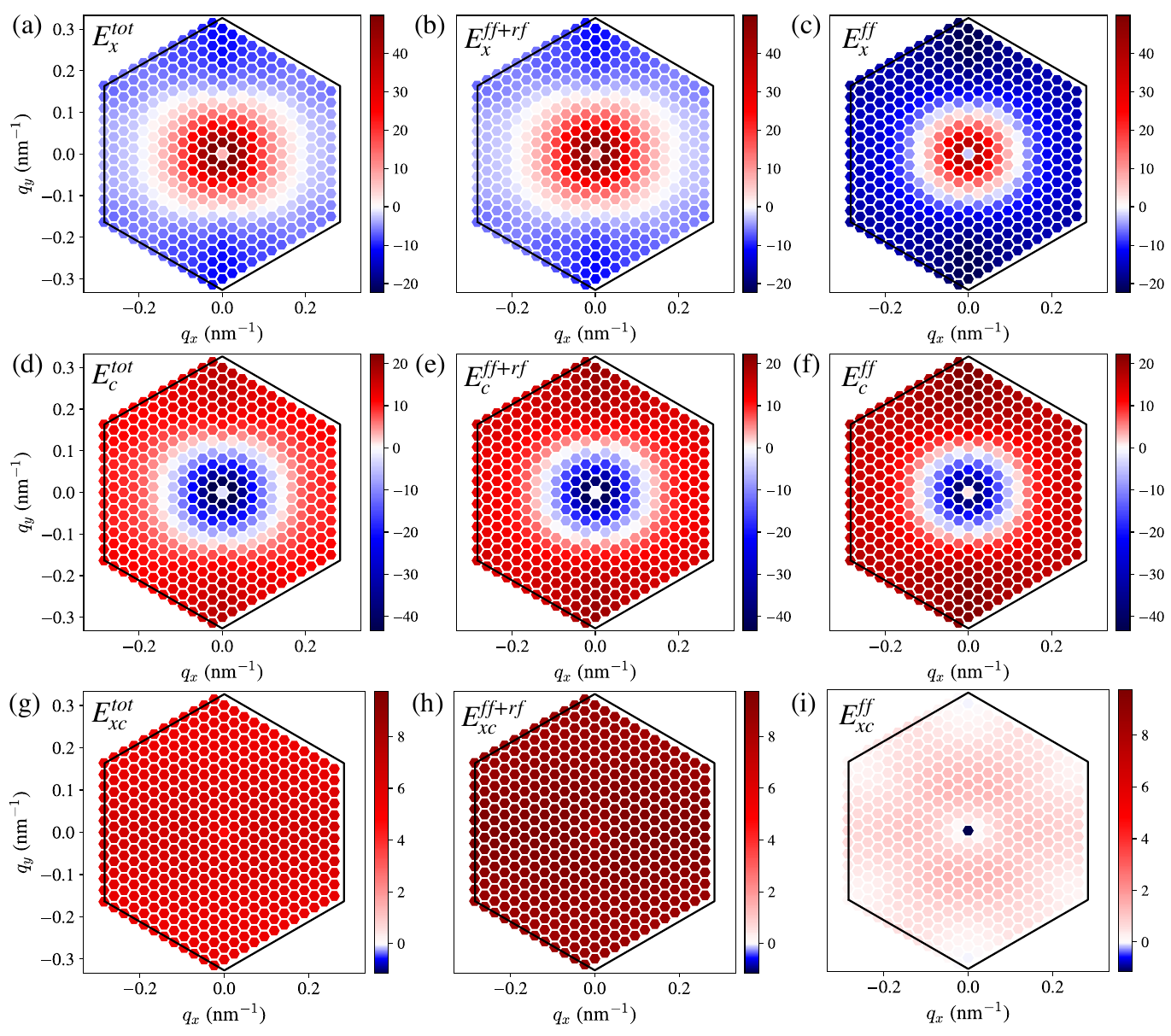}
  \vspace{-10pt}
  \caption{\small Energies versus $\mb{q}$ of the polarized state  at $\nu=-1$.
  }
  \label{Fig:E_vs_q_4000}
\end{figure}

\begin{figure}
\centering  
\includegraphics[width=0.8\columnwidth]{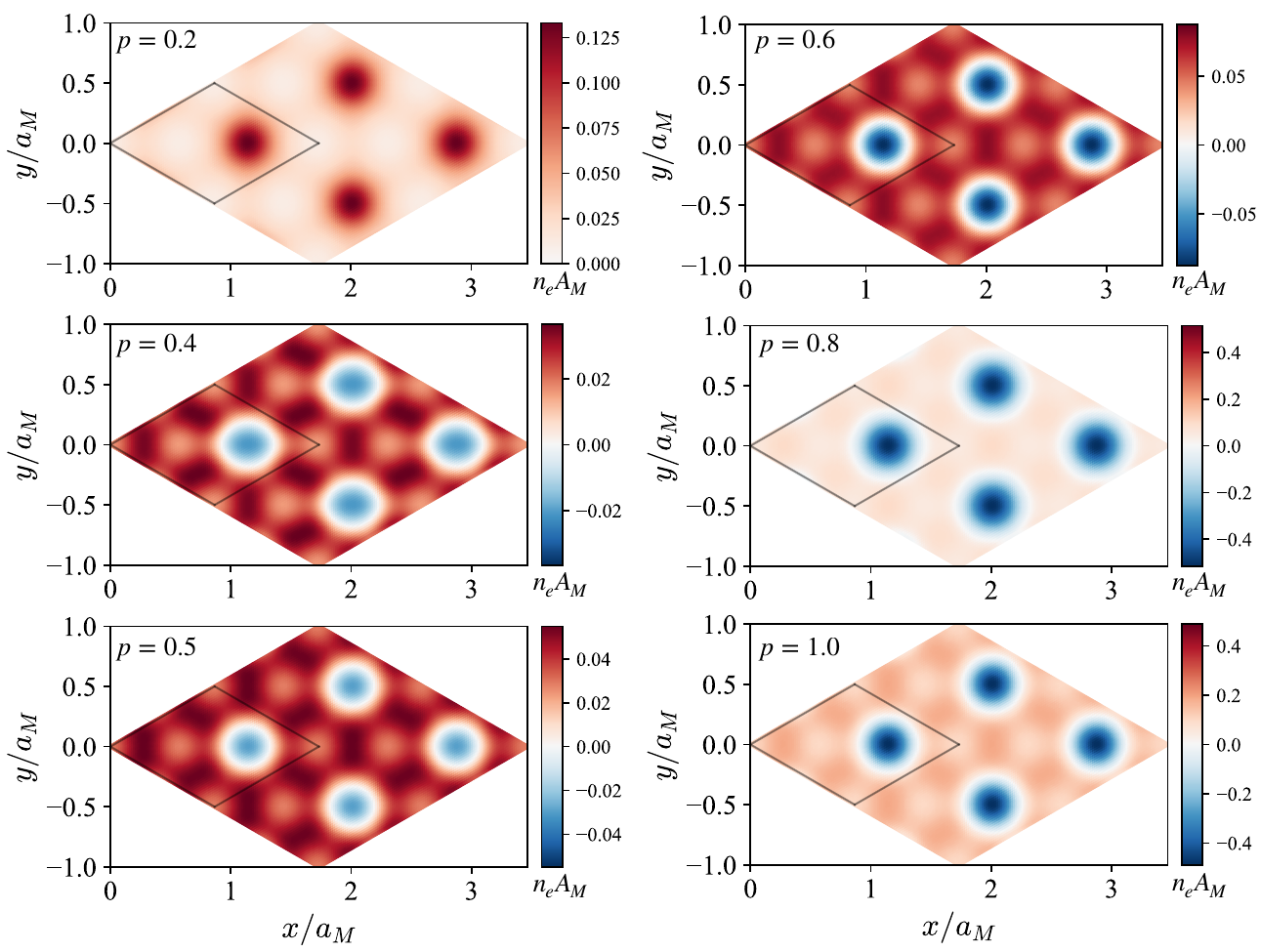}
  \vspace{-10pt}
  \caption{ Density distributions in real-space for $\nu=0$, shown for various polarizations $p = 0.2, 0.4, 0.5, 0.6, 0.8, 1.0$. These correspond to Fig. 3(a) in the main text, where polarization $p$ is defined. Displayed densities account for two flat band contributions and are depicted after subtraction of the $p=0$ baseline. Compared to finite filling cases as shown in Fig.~\ref{Fig:rhor_vs_nu}, densities for $\nu=0$ are at least one order of magnitude smaller and therefore approximately uniform for all $p$.
  Density is in unit of $n_e A_{\rm M}$, where $A_{\rm M}$ is the area of moir\'e unit cell. The grey rhombus outlines the moir\'e unit cell.
  }
  \label{Fig:rhor_vs_p}
\end{figure}

\begin{figure}
\centering  
\includegraphics[width=0.8\columnwidth]{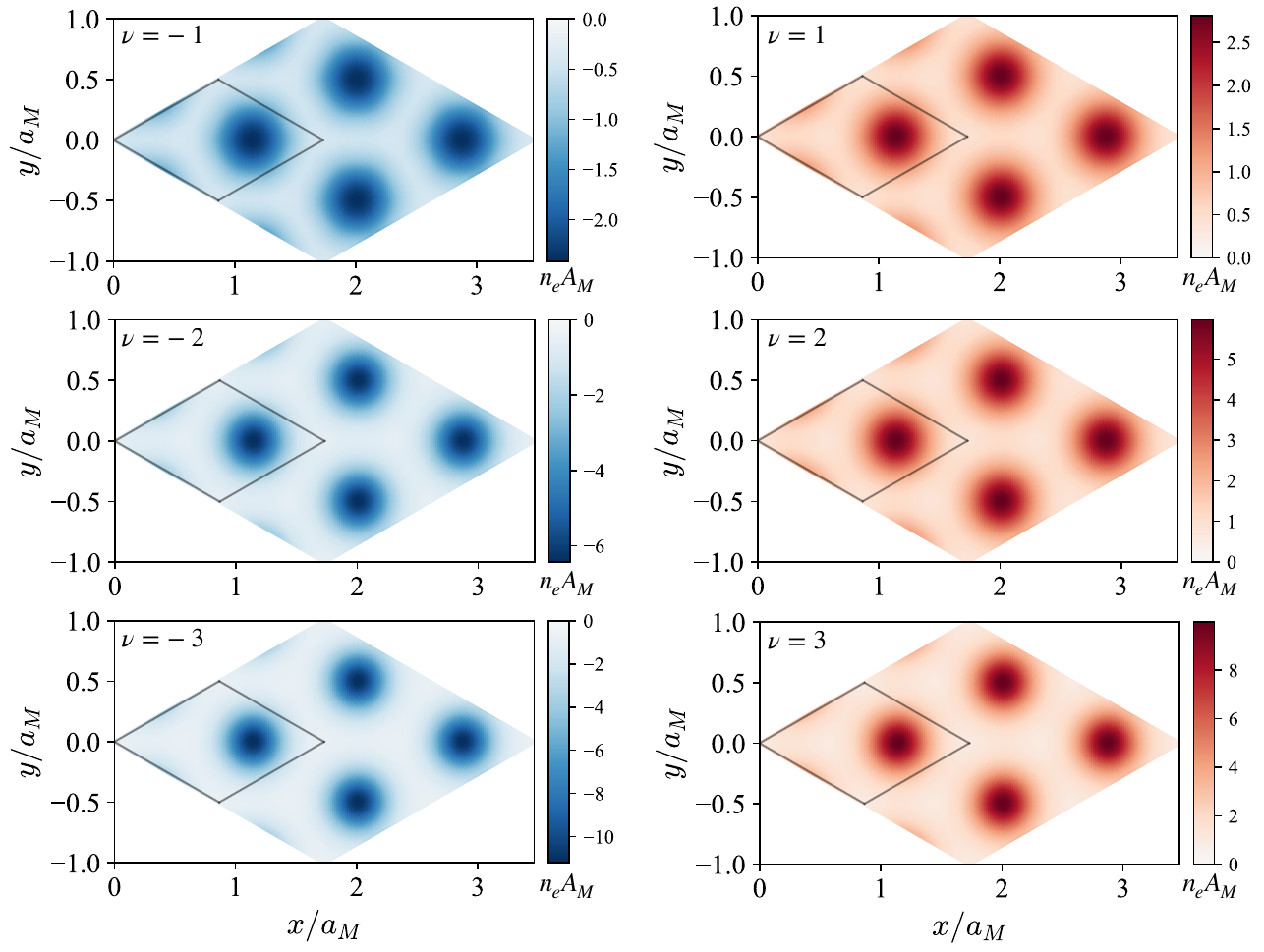}
  \vspace{-10pt}
  \caption{ Density distributions in real-space for the paramagnetic state, as a comparison to Fig.~\ref{Fig:rhor_vs_p}, displayed at different fillings $\nu = \pm 1, \pm 2, \pm 3$. 
  }
  \label{Fig:rhor_vs_nu}
\end{figure}

\begin{figure}
\includegraphics[width=0.4\columnwidth]{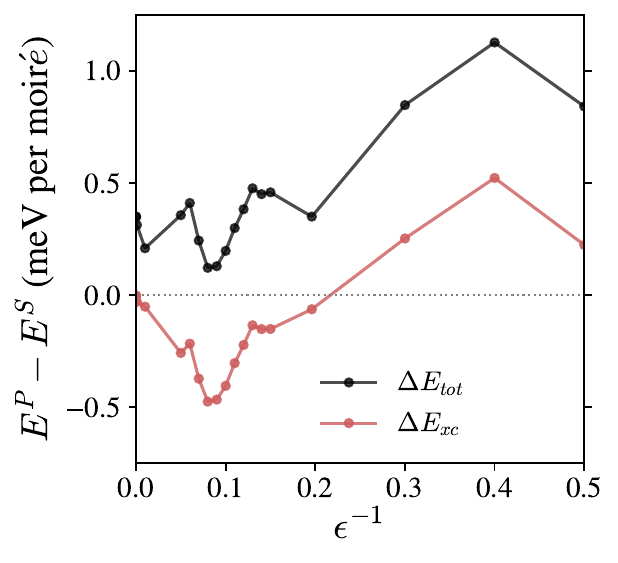}
  \vspace{-10pt}
  \caption{\small The energy difference, defined as $\Delta E = E^P - E^S$, between flavor fully polarized state (P) and flavor paramagnetic state (S) at $\nu=-1$ as a function of Coulomb interaction strength $\epsilon^{-1}$. $\Delta E_{tot}$ and $\Delta E_{xc}$ are energy difference of RPA total and xc energies respectively.
  }
  \label{Fig:E_wrt_EpInv}
\end{figure}

\begin{table*}[t]
\resizebox{\dimexpr \columnwidth}{!}{
\begin{tabular}{ |c|c|c|c|c|c|c||c|c|c|c|c|c|c| }
\hline
$\nu$ & ($\nu_1$, $\nu_2$, $\nu_3$, $\nu_4$) & $E_0$ & $E_x$ & $E_c$ & $E_{xc}$ & $E_{tot}$ & $\nu$ & ($\nu_1$, $\nu_2$, $\nu_3$, $\nu_4$) & $E_0$ & $E_x$ & $E_c$ & $E_{xc}$ & $E_{tot}$ \\
\hline
\multirow{5}{2em}{\centering -1.0} & (-1/4, -1/4, -1/4, -1/4) & -4.119 & 25.398 & -21.363 & 4.034 & -0.084 & \multirow{5}{2em}{\centering 1.0} & (1/4, 1/4, 1/4, 1/4) & 19.664 & 30.622 & -26.822 & 3.800 & 23.464 \\
& (-1/2, -1/4, -1/4, 0) & -3.980 & 23.401 & -19.643 & 3.758 & -0.221 & & (1/2, 1/4, 1/4, 0) & 19.785 & 25.797 & -21.969 & 3.828 & 23.613 \\
& (-1/2, -1/2, 0, 0) & -3.809 & 21.189 & -17.977 & 3.212 & -0.597 & & (1/2, 1/2, 0, 0) & 19.749 & 20.717 & -17.055 & 3.662 & 23.411 \\
& (-3/4, -1/4, 0, 0) & -3.768 & 17.368 & -14.212 & 3.156 & -0.612 & & (3/4, 1/4, 0, 0) & 19.697 & 16.836 & -13.436 & 3.400 & 23.096 \\
& (-1, 0, 0, 0) & -3.500 & 6.145 & -4.249 & \bf{1.897} & \bf{-1.604} & & (1, 0, 0, 0) & 19.796 & 4.637 & -2.015 & \bf{2.622} & \bf{22.418} \\
\hline
\multirow{4}{2em}{\centering -2.0} & (-1/2, -1/2, -1/2, -1/2) & -3.919 & 41.452 & -34.390 & 7.062 & 3.143 & \multirow{4}{2em}{\centering 2.0} & 
(1/2, 1/2, 1/2, 1/2) & 43.469 & 41.064 & -33.315 & 7.749 & 51.218 \\
& (-1, -1/3, -1/3, -1/3) & -3.787 & 30.464 & -23.639 & 6.826 & 3.039 & & 
(1, 1/3, 1/3, 1/3) & 43.534 & 28.905 & -21.296 & 7.609 & 51.144 \\
& (-1, -1/2, -1/2, 0) & -3.106 & 27.850 & -21.761 & 6.089 & 2.983 & & (1, 1/2, 1/2, 0) & 43.903 & 26.511 & -19.296 & 7.215 & 51.118 \\
& (-1, -1, 0, 0) & -2.244 & 14.628 & -10.070 & \bf{4.558} & \bf{2.314} & & (1, 1, 0, 0) & 44.375 & 12.460 & -6.424 & \bf{6.036} & \bf{50.411} \\
\hline
\multirow{5}{2em}{\centering -3.0} & (-3/4, -3/4, -3/4, -3/4) & 2.155 & 47.234 & -38.426 & 8.808 & 10.963 & \multirow{5}{2em}{\centering 3.0} & (3/4, 3/4, 3/4, 3/4) & 72.566 &  45.210 & -33.703 & 11.507 & 84.073 \\
& (-1, -2/3, -2/3, -2/3) & 2.118 & 43.554 & -34.843 & 8.711 & \bf{10.829} & & 
(1, 2/3, 2/3, 2/3) & 72.541 & 41.623 & -30.052 & 11.570 & 84.111 \\
& (-1, -1, -1/2, -1/2) & 2.134 & 36.756 & -27.537 & 9.219 & 11.353 & & (1, 1, 1/2, 1/2) & 72.583 & 35.136 & -23.817 & 11.320 & 83.903 \\
& (-1, -1, -3/4, -1/4) & 2.270 & 34.094 & -25.249 & 8.846 & 11.116 & & (1, 1, 3/4, 1/4) & 72.710 & 32.033 & -20.865 & 11.168 & \bf{83.878} \\
& (-1, -1, -1, 0) & 3.562 & 26.476 & -18.317 & \bf{8.159} & 11.721 & & (1, 1, 1, 0) & 73.749 & 22.621 & -12.401 & \bf{10.219} & 83.968 \\
\hline
\end{tabular}}
\caption{
\label{tab:energy_woC2T}
Same as in Table~\ref{tab:energy} but with broken $C_2T$ by using a massive Dirac Hamiltonian in the BM model. The mass terms in top and bottom layers are both $10$ meV.
}
\end{table*}

\section{Magnetic Anisotropy in SU(4) Ferromagnets}

MATBG has $SU(4)$ ferromagnetism because of its four degenerate spin-valley flavors.
In the continuum model we employ the $SU(4)$ symmetry is reduced to $SU(2) \times SU(2) \times U(1)$
by the difference between two valley projected band Hamiltonians, which contribute explicitly to 
the model's $SU(4)$ magnetic anisotropy.  Valley-exchange and spin-orbit interactions,
which we neglect, also contribute to magnetic anisotropy.  In our RPA approach, we neglect 
anisotropy by focusing only on flavor-dependent filling factors, the eigenvalues of the spin-valley
density matrix.  Indeed our explicit calculations assume that the ferromagnet's density matrix remains 
diagonal in the spin-valley representation because total valley occupation number remains a good quantum
number, simplifying the use of a coupling-constant integral representation of the energy.
There is in fact theoretical\cite{khalaf2021charged,kwan2021kekule,ledwith2021strong,shavit2021theory,shavit2023strain}
and experimental \cite{oh2021evidence,hong2022detecting,yu2022spin,nuckolls2023quantum} work that
intervalley coherence, which breaks valley number symmetry, is present in many MATBG ferromagnets.
Our calculations make no effort to distinguish between ferromagnets that differ only in the 
orientation of the spin-valley magnetization and not in its magnitude as characterized 
by the differences between spin-valley density-matrix eigenvalues.  There is evidence in
recent experiments \cite{nuckolls2023quantum}, in
the form of the presence of sample-specific domain walls and vortices in various partial 
order parameters, that as in most conventional ferromagnets the energy scale associated with anisotropy 
is smaller than the energy scale associated with ordering.  

\end{document}